\documentclass[runningheads]{tlp}

\lefttitle{xxxx}
\jnlPage{1}{8}
\jnlDoiYr{2021}
\doival{10.1017/xxxxx}

\usepackage{pdflscape} %
\usepackage{afterpage}
\usepackage{graphicx}
\usepackage{subcaption}
\DeclareCaptionLabelFormat{Anumber}{1#2}
\graphicspath{ {./plots/} }
\usepackage{changepage}
\usepackage{algorithm}
\usepackage[noend]{algpseudocode}
\usepackage{adjustbox}
\usepackage{ amssymb }
\usepackage{ amsmath }
\usepackage{stackengine}
\usepackage{tablefootnote}
\usepackage[main=english]{babel}
\usepackage{epigraph}
\usepackage{xspace}
\usepackage[utf8]{inputenc}
\usepackage{listings}
\usepackage{color}

\usepackage[
table,
xcdraw,
dvipsnames]{xcolor}
\usepackage{mdwmath}
\usepackage{pdfpages}
\usepackage{fancybox}
\usepackage{appendix}
\usepackage{bookmark}
\usepackage{enumitem}
\usepackage{setspace}
\usepackage[acronym]{glossaries}
\usepackage[T1]{fontenc}
\usepackage{siunitx}

\hypersetup{hidelinks,pageanchor=true,colorlinks,citecolor=Fuchsia,urlcolor=black,linkcolor=Cerulean}

\graphicspath{{./graphics/}}

\DeclareGraphicsExtensions{.pdf,.jpeg,.jpg,.png}

\definecolor{LightCyan}{rgb}{0,0,0}
\definecolor{Cerulean}{rgb}{0,0,0}
\definecolor{Fuchsia}{rgb}{0,0,0}

\newcommand{\zerodisplayskips}{%
\setlength{\abovedisplayskip}{3pt}%
\setlength{\belowdisplayskip}{3pt}%
\setlength{\abovedisplayshortskip}{0pt}%
\setlength{\belowdisplayshortskip}{0pt}}%
\appto{\normalsize}{\zerodisplayskips}
\appto{\small}{\zerodisplayskips}
\appto{\footnotesize}{\zerodisplayskips}
\usepackage{float}
\newfloat{algorithm}{t}{lop}

\definecolor{mGreen}{rgb}{0,0.6,0}
\definecolor{mGray}{rgb}{0.5,0.5,0.5}
\definecolor{mPurple}{rgb}{0.58,0,0.82}
\definecolor{backgroundColour}{rgb}{0.95,0.95,0.92}

\usepackage{listings}

\definecolor{ciaoframe}     {rgb}{  0,    0,  0.3}
\definecolor{ciaostring}    {rgb}{0.6, 0.46, 0.33}
\definecolor{ciaooperators} {rgb}{0.1, 0.15,  0.6}
\definecolor{ciaokeywords}  {rgb}{0.1, 0.15,  0.6}
\definecolor{ciaoassertions}{rgb}{0.1, 0.15,  0.6}
\definecolor{ciaotrust}     {RGB}{200, 130,     0}
\definecolor{ciaocheck}     {rgb}{0.1, 0.2,   0.8}
\definecolor{ciaochecked}   {rgb}{0.2, 0.34,  0.1}
\definecolor{ciaotrue}      {rgb}{0.2, 0.34,  0.1}
\definecolor{ciaofalse}     {rgb}{0.6,  0.0, 0.09}
\definecolor{ciaoprops}     {rgb}{0.1,  0.2,  0.8}
\definecolor{ciaocomment}   {rgb}{0.8,  0.3,  0.3}

\newcommand{\prettylstciao}[0]{
\lstset{language=Prolog,
  frameround=fttt,
  frame=ltrb,
  rulecolor=\color{ciaoframe},
  numbers=left,numberstyle=\tiny,stepnumber=1,numbersep=8pt,
  tabsize=4,
  breaklines=true,breakatwhitespace=true,
  basicstyle=\scriptsize\ttfamily, %
  showlines=true,
  showspaces=false,
  showtabs=false,
  escapechar=@,
  escapeinside={~~},
  commentstyle=\color{ciaocomment},
  stringstyle=\color{ciaostring},
  showstringspaces=false,
  deletekeywords={true}, %
  keywordstyle={\color{ciaooperators}\bfseries}, %
  classoffset=1, %
        otherkeywords={>,<,>=,=<,.,;,-,!,=,*,\&,+,:-,[,],|,->,:,:=,\#},
        keywordstyle={\color{ciaokeywords}\bfseries},
  classoffset=2,
       morekeywords={module,use_module,dynamic,export,import,multifile,impl_defined,trait,impl,mode},
       keywordstyle={\color{ciaokeywords}\bfseries},
       morekeywords={pred,prop,calls,success,comp},
       keywordstyle={\color{ciaoassertions}\bfseries},
  classoffset=4,
       morekeywords={trust,trust_default,entry},
       keywordstyle={\color{ciaotrust}\bfseries},
  classoffset=5,
       morekeywords={check},
       keywordstyle={\color{ciaocheck}\bfseries},
  classoffset=6,
       morekeywords={checked},
       keywordstyle={\color{ciaochecked}\bfseries},
  classoffset=7,
       morekeywords={true},
       keywordstyle={\color{ciaotrue}\bfseries},
  classoffset=8,
       morekeywords={false},
       keywordstyle={\color{ciaofalse}\bfseries},
  classoffset=9,
       morekeywords={even,nat,str,int,flt,atm,term,num,var,list,ground,mshare,
                    rsize,cardinality,not_fails,exp,cost,costb,steps_ub,steps_lb,
                    size_ub,size_lb,covered,mut_exclusive,head_cost,literal_cost,
                    is_det,non_det,length,terminates,steps_o,resource,socket,seff,string,
                    lowercase
       },
       keywordstyle={\color{ciaoprops}\bfseries},
  classoffset=0, %
}}

 \usepackage{tikz}
\usetikzlibrary{shapes.geometric}
\usetikzlibrary{matrix}
\usetikzlibrary{shapes.gates.logic.US}
\usetikzlibrary{circuits}
\usetikzlibrary{positioning,shapes,arrows,fit,shadows,calc,matrix}
\usetikzlibrary{backgrounds}

\usetikzlibrary{arrows}
\usetikzlibrary{arrows.meta}

\usepackage{forest}

\newcommand{\pneck}{\: \mbox{{:}{-}} \:}

\newcommand{\ciao}{\texttt{Ciao}\xspace}
\newcommand{\ciaopp}{\texttt{CiaoPP}\xspace}
\newcommand{\lpdoc}{\texttt{LPdoc}\xspace}

\makeatletter
\csdef{forest@compute@node@boundary@rounded rectangle}{%
  \forest@mt{east}%
  \forest@lt{north east}%
  \forest@lt{north}%
  \forest@lt{north west}%
  \forest@lt{west}%
  \forest@lt{south west}%
  \forest@lt{south}%
  \forest@lt{south east}%
  \forest@lt{east}%
}
\makeatother

\usepackage{wrapfig}

\DeclareOldFontCommand{\tt}{\ttfamily}{\mathtt}

\definecolor{lightred}{rgb}{0.8,  0.3,  0.3} 
\definecolor{forestgreen}{rgb}{34, 139, 34}
\usepackage[textsize=scriptsize,textwidth=1.7cm]{todonotes}

\usepackage{pifont}%

\newcommand{\literal}{\txt{Literal}}
\newcommand{\plai}{PLAI\xspace}
\newcommand{\goal}{\txt{Goal}}
\newcommand{\call}{\txt{Call}}
\renewcommand{\prime}{\txt{Prime}} 
\newcommand{\head}{\txt{Head}}
\newcommand{\proj}{\txt{Proj}}
\newcommand{\body}{\txt{Body}}
\newcommand{\bd}[1]{\txt{B}\ensuremath{_{#1}}}
\newcommand{\entry}{\txt{Entry}}
\renewcommand{\succ}{\txt{Success}}
\newcommand{\exit}{\txt{Exit}}
\newcommand{\term}[1]{\txt{T}\ensuremath{_{#1}}\xspace}
\newcommand{\builtin}{\txt{abstractLiteral}}
\newcommand{\asub}[1]{\txt{Abs}\ensuremath{_{#1}}}

\newcommand{\vars}{\txt{vars}}
\newcommand{\augment}{\txt{augment}}
\newcommand{\lub}{\txt{computeLub}}

\newcommand{\project}{\txt{project}}
\newcommand{\extend}{\txt{extend}}
\newcommand{\calltoentry}{\txt{callToEntry}}
\newcommand{\exittoprime}{\txt{exitToPrime}}

\newcommand{\entrytoexit}{\txt{entryToExit}}

\newcommand{\dom}{\txt{dom}}

\newcommand{\topmost}{\txt{topmost}}

\newcommand{\txt}[1]{\texttt{#1}}

\newcommand{\mtit}[1]{\ensuremath{\mathit{#1}}}

\renewcommand{\paragraph}[1]{\noindent\textbf{\emph{#1}}}

\usepackage{booktabs} %

\begin{document}

\title[Abstract Environment Trimming]{Abstract Environment Trimming\thanks{
    Partially funded by MICINN projects
    PID2019-108528RB-C21 \emph{ProCode},
    TED2021-132464B-I00 \emph{PRODIGY},
    and by the Tezos foundation. \\
    We also thank the anonymous reviewers for their very useful feedback.
  }}

\begin{authgrp}
\author{\sn{Daniel} \gn{Jurjo-Rivas}$^\dagger$ \ \ \ \sn{Jose F.} \gn{Morales}$^\dagger$}
\author{\sn{Pedro} \gn{L\'opez-Garc\'ia}$^{\dagger\dagger}$  \ \ \ \sn{Manuel V.} \gn{Hermenegildo}$^\dagger$}
\affiliation{\ }
\affiliation{$^\dagger$Universidad Polit\'ecnica de Madrid (UPM) \& IMDEA Software Institute, Spain}
\affiliation{$^{\dagger\dagger}$Spanish Council for Scientific Research \& IMDEA Software Institute, Spain}
\end{authgrp}

\history{\sub{xx xx xxxx;} \rev{xx xx xxxx;} \acc{xx xx xxxx}}

\maketitle
\begin{abstract}
  Variable sharing is a fundamental property in the static analysis of
  logic programs, since it is instrumental for ensuring correctness
  and increasing precision while inferring many useful program
  properties. Such properties include modes, determinacy, non-failure,
  cost, etc.
  This has motivated significant work on developing abstract
  domains to improve the precision and performance of sharing analyses.
  Much of this work has centered around the family of \emph{set-sharing}
  domains, because of the high precision they offer. 
  However, this comes at a price:
  their scalability
  to a wide set of realistic programs remains challenging and this
  hinders their wider adoption.
  In this work, rather than defining new
  sharing abstract domains, we focus instead on developing techniques
  which can be incorporated in the analyzers to address aspects that
  are known to affect the efficiency of these domains, such as the
  number of variables, without affecting precision.  These techniques
  are inspired in others used in the context of compiler
  optimizations, such as expression reassociation and variable
  trimming. We present several such techniques and provide an
  extensive experimental evaluation of over 1100 program modules taken
  from both production code and classical benchmarks. This includes
  the Spectector cache analyzer, the s(CASP) system, 
  the libraries of the \ciao system, the \lpdoc
  documenter, the \plai analyzer itself, etc.
  The experimental results are quite encouraging:
  we have obtained significant speed-ups, 
  and, more importantly, the
  number of modules that require a timeout was cut in half. As a
  result, many more programs can be analyzed precisely in reasonable
  times.
\end{abstract}
\reversemarginpar

\vspace*{-7mm}
\section{\textbf{Introduction}}
Abstract Interpretation~\citep{Cousot77-short} 
allows
constructing sound program analysis tools
that can extract properties of a program by safely approximating its semantics.
Abstract interpretation-based analysis was proved practical and
effective in the context of (Constraint) Logic Programming
((C)LP)~\citep{anconsall-acm-short,softpe,LeCharlier94:toplas-short,mcctr-fixpt-short,ai-jlp-short,vanroy-naclp90-short,pracabsin-short},
which was also one of its very first application
areas~\citep{DBLP:journals/annals/GiacobazziR22}, and the techniques
developed for CLP have also proved useful in the analysis and
verification of other programming languages by using semantic
translation into Constraint Horn Clauses 
(CHCs)~\citep{anal-peval-horn-verif-2021-tplp-short,HGScam06-short,decomp-oo-prolog-lopstr07-short}.
In the context of static analysis of (C)LP programs, variable sharing
soon emerged as a fundamental property, which has led to very active
and continuous development of variable sharing analysis domains.
Sharing proved immediately necessary for ensuring correctness and
precision while inferring most other useful program properties such as
modes, determinacy, non-failure,
and cost, among others.  In fact, some early LP analyses were actually
incorrect because variable sharing was not taken into account.
Sharing analyses have also proven fundamental in the optimization of
unification ~\citep{sonder86} and 
in automatic (and-)parallelization 
~\citep{sinsi-jlp,nsicond-sas94,effofai-toplas-short,dep-par-iclp,consind-toplas}. E.g.,
in parallelization it is crucial to precisely detect whether two
variables are independent (i.e., they do not \emph{share}), a variable
is ground, etc.  Sharing has also been used as the basis for more
complex analyses for related properties such as linearity, paths, or
freeness~\citep{pathsh-iclp94,DBLP:journals/tplp/AmatoS14,DBLP:journals/mscs/AmatoMS22,BruynoogheCodish93,freeness-iclp91-medium}.
Furthermore, beyond (C)LP, sharing analysis is directly related to
\emph{aliasing} and \emph{points-to} analyses in imperative
programming, widely used in the context of languages with pointers and
dynamic
memory~\citep{AikenFKT03,DBLP:conf/pldi/LandiR92a,steensgaard96pointsto,Datalog:DOOP-short,resources-bytecode09,rountev01pointsto,DBLP:conf/sas/WhaleyL02},
sometimes applying directly domains stemming from
(C)LP~\citep{zanardini:fieldsharing,shnltau-vmcai08}.  In fact, (C)LP
sharing analyses constituted some of the very first abstract
interpretation-based pointer aliasing analyses for any programming
language.

In this paper we concentrate on a popular abstract domain for variable
sharing analysis: \emph{set-sharing}
analysis~\citep{jacobs89-short,abs-int-naclp89-short}. This domain
captures which \emph{sets} of program variables share run-time variables,
encoding this information in \emph{sharing sets}. 
For example, assume $X$, $Y$, and $Z$ are the syntactic program
variables that we need to track,
and consider the substitution (run-time store)
$\{X/f(M, K, a), Y/g(b, M), Z/g(a,b)\}$. 
This substitution
is abstracted to the sharing set
$\{\{X\}, \{X, Y\}\}$,
where $\{X, Y\}$ represents that there is (or, more precisely, may be) one
or more variables shared in the terms
to which $X$ and $Y$ are bound,
and $\{X\}$ represents that there is one or more variables that appear
only in $X$.
Additionally, $Z$ not appearing in any set means that it contains no
variables, i.e.,
$Z$ is bound to a ground term.
Set-sharing encodes not only variable independence, i.e., the fact
that substitutions that affect a given variable will not affect
another one, but also groundness, grounding dependencies (e.g., the
fact that if $Y$ becomes ground then $X$ becomes ground, but not the
other way around), independence relationships, etc.
This representation richness does come, however, at a price: the
scalability of set-sharing domains to a wide set of realistic programs
is challenging, since the number of sharing sets carried by the
abstraction
can be exponential in the number of variables 
of the clause being analyzed.
This has prompted much work in improvements and alternative
representations of set-sharing abstractions, with the objective of
improving performance while maintaining precision as much as possible.
In contrast, in this work, rather than defining new sharing abstract
domains or modifying existing ones, we concentrate 
on
developing techniques that can be incorporated in the analysis
framework to address the root causes of the performance issues faced
by the set-sharing domains, such as the number of variables, without
affecting precision.
We draw inspiration from techniques used in the context of
compiler optimizations,
which significantly reduce the number of
variables presented in the abstractions.

The rest of the paper is structured as follows: First,
Section~\ref{sec:prelim} provides the necessary background, covering
set-sharing abstract domains and top-down
analysis. Section~\ref{sec:reassociation-and-trimming} 
presents 
our approach, offering first a program
transformation that can provide an optimal solution; and second,
an alternative solution based on \emph{variable trimming} 
that can be applied during analysis without
modifying the program. Section~\ref{sec:experiment} reports our
experimental evaluation 
and finally, Section~\ref{sec:conclusions} summarizes our conclusions
and discusses
some lines of future work.
\vspace*{-5mm}
\section{\textbf{Notation and Preliminaries}}
\label{sec:prelim}
We represent variables by uppercase letters (for example: $X$, $Y$,
$Z$, etc.) and atoms by lowercase letters (for example: $a$, $b$, $c$,
etc.).  $\mathcal{P}(S)$ denotes the powerset of set $S$ and
$\mathcal{P}^0(S)$ the \emph{proper} powerset of set $S$,
i.e., $\mathcal{P}^0(S)=\mathcal{P}(S)\backslash \{\emptyset\}$. Given
a term \term{}, 
\vars(\term{}) denotes the set of its variables. 
A Constraint Logic Program (CLP) is a set of \emph{clauses} of the
form $H \pneck A_{1}, \ldots, A_{n}$ where $A_{1}, \ldots, A_{n}$ are
\emph{literals} that form the \emph{body} and H is a
positive literal said to be the \emph{head} of the clause.
A \emph{substitution} is a set $\theta = \{V_1/t_1, \ldots, V_n/t_n\}$
with $V_i$ distinct variables and $t_i$ terms.
We say that $t_i$ is
the \emph{value} of $V_i$ in $\theta$. The set $\{V_1,\ldots,V_n\}$ is
the \emph{domain} of $\theta$ 
(\dom($\theta$)).
\\
The main idea behind \emph{Abstract Interpretation}~\citep{Cousot77-short} is
to interpret the program over a special, abstract domain whose
elements are finite representations of possibly infinite sets of
actual substitutions in the concrete domain. We 
denote the concrete domain as $D_{\gamma}$, the abstract domain as
$D_{\alpha}$, 
and
the functions that relate sets of concrete substitutions with
abstract substitutions as the \emph{abstraction} function $\alpha : D_{\gamma} \rightarrow D_{\alpha}$ and the \emph{concretization} function $\gamma : D_{\alpha} \rightarrow D_{\gamma}$.
The concrete domain is typically a complete lattice with the set
inclusion order which induces an ordering relation in the abstract
domain 
represented by $\sqsubseteq$. Under this relation the abstract domain
is usually a complete lattice and
$(D_{\gamma}, \alpha, D_{\alpha}, \gamma)$ is a Galois
insertion/connection~\citep{Cousot77-short}.
Several frameworks
for abstract interpretation exist; this work focuses on
\emph{top-down frameworks},
discussed in 
Section~\ref{sec:plai}.

\medskip
\paragraph{\textbf{Set-Sharing Analyses.}}\label{sec:sharing}
As mentioned in the introduction, in static analysis of logic
programs, tracking of variables shared among terms is essential.
A set of program variables
$V_1,\ldots, V_m$ \emph{share} if, in some execution of the program,
they may respectively be bound to terms $\term{1},\ldots,\term{m}$
such that
$vars(\term{1})\cap\ldots \cap vars(\term{m}) \neq \emptyset$.  One of
the most accurate abstract domains defined for tracking sharing
information is
\emph{set-sharing}~\citep{jacobs89-short,abs-int-naclp89-short}.
This domain
captures whether
at run-time there are variables sharing, 
condensing this information in a concise set representation.
As an example, consider
program variables \txt{X}, \txt{Y},
\txt{Z}, \txt{W}, \txt{T}, and assume they are bound at run-time as
follows:
$\theta=\{\txt{X/f(M,a), Y=g(b,M,c), Z/[1,M,3], W/g(b), T/h(K,m)}\}$.
This substitution (run-time state) is represented in the
\emph{set-sharing} domain by the \emph{sharing} abstraction
$\{\{\txt{X,Y,Z}\},\{\txt{T}\}\}$. The first element,
$\{\txt{X,Y,Z}\}$, represents the fact that there is (at least one)
variable (i.e., \txt{M}) that appears in all of \txt{X}, \txt{Y},
\txt{Z}, but not in \txt{T} or \txt{W}; and the second element,
$\{\txt{T}\}$, represents that there is (at least one) variable that appears in
\txt{T} (i.e., \txt{K}) but not in any of the others.  We say that
\txt{X}, \txt{Y} and \txt{Z} ``share'', and that \txt{T} does not
``share'' with \txt{X}, \txt{Y}, or \txt{Z}. The fact that \txt{W}
does not appear in any set means it contains no variables and thus,
it is ground.  This representation also captures that there are
no other variables in \txt{X}, \txt{Y}, or \txt{Z} in addition to the
one(s) they share, which has important implications with respect to
grounding: after a program statement that grounds one of them (e.g.,
\txt{Z=[1,2,3]}), we know both \txt{X} and \txt{Y} will also be
grounded. However, grounding \txt{T} does not ground any of \txt{X},
\txt{Y}, or \txt{Z}.
Other abstract domains have also been studied,
notably the \emph{pair-sharing},
which keeps
track of \emph{pairs} of variables that share.  E.g., for the example above,
its pair-sharing abstraction is:
$\{(\txt{X}, \txt{Y}), (\txt{Y},\txt{Z}), (\txt{X}, \txt{Z})\}$. 
The intricacies of the relation and trade-offs between set-sharing and
pair-sharing are beyond the scope of this paper; the reader is
referred to, for example,~\cite{BagnaraHZ97SAS}
and~\cite{setsh-flops04} for further discussion of this topic.
However, our subject of study in this work is set-sharing analyses. 
  
The 
set-sharing domain has attracted a lot of attention in the literature
and has been enhanced in different ways
and extended with other kinds of information such as
\emph{freeness} or \emph{linearity}~\citep{%
  freeness-iclp91-medium,bruynooghe:composite,file-sharing-tr,pathsh-iclp94,
  codish96freeness,Fecht-plilp96,Zaffanella99widening-short,
  HillZB04TPLP,
  shcliques-padl06-short,
  negsharing-iclp08,DBLP:journals/tplp/AmatoS09,DBLP:journals/mscs/AmatoMS22}.

However, the set representation, which allows the sharing domain to
offer high precision, is also one of the reasons why this family of
domains presents scalability challenges.  A set-sharing abstraction is
presented as a set of sets, each of them capturing a \emph{possible}
sharing that occurs at runtime. In cases where there is not much (or
any) sharing information at runtime, more (or all the) sharing sets
are possible.  Given a set of variables appearing in a program being
analyzed ($V$), the size of a set-sharing abstraction has an upper
bound given by the abstraction which captures all the possible
non-empty sharing sets ($\mathcal{P}^0(V)$).  Thus, the size of an
abstraction is, in the worst case, exponential in terms of the number
of variables that appear in the program.  To overcome these problems,
various representations have been proposed, such as collapsing subsets
of the abstraction into ``cliques'' (sets of variables that represent
the proper powerset of those variables). These representations allow
for a reduction in the size of the sharing abstraction and can improve
performance, even more so when equipped with widenings (albeit then at
the cost of losing precision)~\citep{Zaffanella99widening-short,
  shcliques-padl06-short}. For example, the set-sharing abstraction
$\{\{\txt{X}\},\{\txt{X,Y}\},\{\txt{Y}\},\{\txt{Z}\}\}$ can be
represented using cliques as the tuple
$(\{\{\txt{X,Y}\}\}, \{\{\txt{Z}\}\})$ where the clique
$\{\txt{X,Y}\}$
represents $\mathcal{P}^0(\{\txt{X,Y}\})$.

\medskip 
\paragraph{\textbf{The \plai Top-Down Analyzer.}}\label{sec:plai}
\emph{Top-down} analyses are a family of static analyses that build an
\emph{analysis graph} starting from a series of program \emph{entry
  points}. This approach was first used in analyzers such as MA3 and
Ms~\citep{pracabsin-short}, and matured in the \plai
analyzer~\citep{mcctr-fixpt-short,ai-jlp-short} using an optimized
fixpoint algorithm now also referred to as the \emph{top-down algorithm}
or \emph{solver}. This algorithm was later 
 applied to the analysis of CLP/CHCs~\citep{anconsall-acm-short} and
imperative
programs~\citep{anal-peval-horn-verif-2021-tplp-short,HGScam06-short,decomp-oo-prolog-lopstr07-short,fixpt-javabytecode-bytecode07-short},
and used in analyzers such as GAIA~\citep{LeCharlier94:toplas-short},
the CLP($\cal R$) analyzer~\citep{softpe}, or
Goblint~\citep{seidl3improv,tdSeidlA2I}.

The graph constructed by the \plai algorithm during analysis 
is a finite, abstract object whose concretization approximates the
(possibly infinite) set of (possibly infinite) maximal AND-trees of
the concrete semantics.
This 
approach separates the abstraction of the structure of
the concrete trees (the paths through the program) from the abstraction of the
\emph{substitutions} at the nodes in those concrete trees (the program
states in those paths). The first abstraction, $T_{\alpha}$, is 
typically built-in, 
as an abstract domain of \emph{analysis graphs}. 
The framework is \emph{parametric} on 
a second abstract domain, $D_{\alpha}$, 
whose elements appear as 
labels in the nodes of the analysis 
graph. 
A more
detailed recent
discussion can be found
in~\citep{anal-peval-horn-verif-2021-tplp-short}.
Assume we are analyzing a literal \goal~in the body of some clause in
the program, and that \head$\,\pneck\,$\body~is a clause in a predicate whose
head unifies with \goal. 
Assume also that the substitution affecting \goal~at the time of this
call is approximated by the abstract substitution \call~such that
 $\vars(\goal)$$\subseteq$$\dom(\call)$ and
$\vars(\call)\cap(\vars(\head)\cup\vars(\body))=\emptyset$.
The success (exit state) of \goal~after having executed the above clause is
represented by the abstract substitution \succ~given by:
\vspace*{-0.7mm}
\begin{align*}
  \succ  &= \extend(\call,\goal,\prime) \\[-1.2mm]
  \prime &= \exittoprime(\project(\vars(\head), \exit), \head, \goal) \\[-1.2mm]
  \exit  &= \entrytoexit(\entry, \head, \body) \\[-1.2mm]
  \entry &= \augment(\vars(\body)\backslash\vars(\head), \calltoentry(\proj, \goal, \head)) \\[-1.2mm]
  \proj  &= \project(\vars(\goal), \call) 
\end{align*}

\vspace*{-1mm}
  As an example, let \goal\,$=\txt{p(A,f(B),E)}$,
  \call\,$=\{\{\txt{A}\},\{\txt{B,C}\}, \{\txt{A,C,D}\}\}$
  (notice that \txt{E} is ground, since it does not appear in any sharing set)
  and \head$\,\pneck\,$\body{} be the clause
  \txt{p(f(X),f(Y),Z)}$\,\pneck\,$\txt{[X|T1]=[X,Y|T2]}, whose
  \head~unifies with \goal. The success abstraction is computed as
  follows:
\vspace*{-1mm}
\begin{itemize}  
\item[i)] \label{step:project}
  First, the abstraction \call{} is \emph{projected} onto the
  variables in \goal{} by means of the \project\ function, obtaining
  \proj\,$=\{\{\txt{A}\}, \{\txt{B}\}\}$.
\item[ii)] Then, \calltoentry(\proj, \goal, \head) yields an abstract
  substitution that represents the unification
  \txt{p(A,f(B),E)}=\txt{p(f(X),f(Y),Z)} (i.e., \goal=\head) in the context
  represented by
  \proj. 
  In our example, such abstraction is $\{\{\txt{X}\}, \{\txt{Y}\}\}$,
  where \txt{Z} is becomes ground because it is unified with \txt{E},
  which is ground.

\item[iii)] Now, the domain of such abstraction is extended with the
  variables in \body{} that do not appear in \head{} (i.e., \txt{T1}
  and \txt{T2}), and the abstraction is updated accordingly by
  including safely approximated information. This is done by operation
  \augment, which returns the \entry{} abstraction $\{\{\txt{X}\},
  \{\txt{Y}\}, \{\txt{T1}\}, \{\txt{T2}\}\}$.

\item[iii)] Then, \body{} is traversed so that each of its literals
  are (recursively) analyzed by procedure \Call{entryToExit}{},
  described in Algorithm~\ref{alg:entry_to_exit}.\\  In our example,
  \Call{entryToExit}{\entry, \head, \body} proceeds as follows:
  \begin{itemize}
  \item[a)] First, the \exit~abstraction is initialized with the
    current \entry~abstraction (Line~\ref{alg:ana-entry2exit-init}),
    and the first literal of the body is selected
    (Line~\ref{alg:ana-entry2exit-loop}), which in this case
    corresponds to the only literal in the body: \txt{[X|T1]=[Z,Y|T2]}.
  \item[b)] The \plai framework proceeds differently depending on the
    kind of literal being analyzed (see
    Lines~\ref{alg:entry2exit:rec-beg}--\ref{alg:entry2exit:builtin-end}).
    Since the literal \txt{[X|T1]=[Z,Y|T2]} is neither a recursive
    call nor a call to a predicate in the analysis scope, the steps in
    Lines~\ref{alg:entry2exit:builtin-beg}--\ref{alg:entry2exit:builtin-end}
    are performed as explained below.
    
  \item[c)] First, the abstraction is projected onto the variables in
    the literal, returning $\{\{\txt{X}\}, \{\txt{Y}\}, \{\txt{T1}\},
    \{\txt{T2}\}\}$) and the operation \builtin~is invoked, which
    captures the abstract behavior of the literal. In our example, it
    performs the abstract unification \txt{[X|T1]=[Z,Y|T2]}. Since
    \txt{Z} is ground, and the unification induces \txt{X=Z}, the
    groundness information is propagated to \txt{X}. Such unification
    also induces \txt{T1=[Y|T2]}, which results in the creation of new
    sharing sets accordingly. The \asub{\txt{exit}} abstraction
    obtained after these operations is $\{\{\txt{Y},\txt{T1}\},
    \{\txt{Y},\txt{T1},\txt{T2}\}, \{\txt{Y},\txt{T2}\}\}$.

  \item[d)] \label{step:extend} Finally, such
    abstraction is used to update the previous exit abstraction by the
    \extend~operation (Line~\ref{alg:entry2exit:builtin-end}),
    yielding \exit=$\{\{\txt{Y},\txt{T1}\},
    \{\txt{Y},\txt{T1},\txt{T2}\}, \{\txt{Y},\txt{T2}\}\}$.
  \end{itemize}
  
\vspace*{-1mm}
\item[iv)] After the execution of \Call{entryToExit}{}, the obtained
  \exit~abstraction is projected over \vars(\head) and
  represented in the context of the variables of \call. This is done
  by operation \exittoprime(\project(\vars(\head), \exit), \head,
  \goal), which captures the effects of the unification
  \head=\goal.
  In our example, it yields \prime=$\{\{\txt{B}\}\}$,
  since the groundness information has been propagated from \txt{X} to
  \txt{A}.
\item[v)] The analysis concludes with the \extend(\call, \prime)
  operation, which \emph{updates} the initial \call~abstraction with
  the new inferred information 
  in the \prime~abstraction,
  obtaining the success abstraction: $\succ=\{\{\txt{B,C}\}\}$, where
  the sharing-set $\{\txt{A,C,D}\}$ has been deleted because \txt{A} is
  ground,
  and such
  information
  is propagated to \txt{D}.
\end{itemize}

As some final remarks, if no predicate head can be unified with the
goal under analysis, a bottom abstraction ($\bot$) is returned
(representing that the exit state is unreachable). If several clauses
are available, all of them are analyzed, and a collection of
\emph{prime} abstractions $\prime_1,\ldots,\prime_m$ is obtained, one
abstraction per clause, where $m$ is the number of clauses. Then, the
success abstraction is computed as
\succ=\extend(\call,\lub(${\prime_1,\ldots,\prime_m}$)), where
\lub\ yields the \emph{least upper bound} of the collection of
abstractions (other operators, including disjunction and
widenings, are possible).

In the \Call{entrytoexit}{} loop
(Lines~\ref{alg:ana-entry2exit-loop}-\ref{alg:entry2exit:builtin-end}),
when the current literal under analysis corresponds to a recursive call
(Lines~\ref{alg:entry2exit:rec-beg}-\ref{alg:entry2exit:rec-end}), the
analyzer computes a fixpoint for the associated call pattern.  Such
call pattern is
determined by the current literal \goal{} and the abstraction \call{}
representing the environment under which \goal{} is executed (this
may require the use of a widening operation to ensure termination).
A detailed discussion on the different fixpoint computation methods is
outside the scope of this work and we believe that it is not necessary
for understanding our approach and contribution. The reader is referred
to, e.g.,~\cite{mcctr-fixpt-short,ai-jlp-short} for more details. 
 
In the case that the current literal is not recursive but corresponds
to a call to a predicate within the analysis scope, the associated
call pattern is analyzed
(Lines~\ref{alg:entry2exit:in-scope-beg}-\ref{alg:entry2exit:in-scope-end})
following steps i) to v)
illustrated above using our example, with the clauses that unify with the literal.

Finally, if the literal to be analyzed does not correspond to any of the two
cases discussed above
and the analysis domain does not know how to abstract it either
(Lines~\ref{alg:entry2exit:builtin-beg}--\ref{alg:entry2exit:builtin-not-know}),
the invocation to the \builtin\,operation returns a \txt{fail}
atom. The analyzer then computes the \emph{top-most} information
for \vars(\literal) by calling the \topmost\,function
(Line~\ref{alg:entry2exit:builtin-topmost}). Then, the original
\call{} abstraction is extended with such top-most information. In our
example, if the abstract domain did not implement how to abstract the
unification \txt{[X|T1]=[Z,Y|T2]}, the top-most abstraction would be
$\mathcal{P}^0(\{\txt{X,T1,T2,Y,Z}\})$. Notice that this can be quite
common when, for example, the analyzer has to process a call to a
predicate which is imported from a library whose source code is not
available, is not in the analysis scope, etc.  In this case, the
top-most abstraction has to be assumed to ensure correctness.

\begin{algorithm}
\vspace*{-0.5mm}
  \caption{A schematic description of \entrytoexit}
  \label{alg:entry_to_exit}
  \begin{algorithmic}[1]
    \Function{\entrytoexit}{$\entry, \head, \body$}\label{alg:ana-entry2exit-beg}
    \State $\exit \gets \entry$ \label{alg:ana-entry2exit-init}
    \For{$\txt{Literal} \in \body$}\label{alg:ana-entry2exit-loop}
    \If{$\Call{recursive-call}{\txt{Literal}, \head}$} \label{alg:entry2exit:rec-beg}
        \State $\exit \gets \Call{compute-fixpoint}{\exit, \literal}$\label{alg:entry2exit:rec-end}
    \ElsIf{$\Call{predicate-in-scope}{\txt{Literal}}$} \label{alg:entry2exit:in-scope-beg}
        \State $\exit \gets \Call{analyze-pred}{\exit, \literal}$ \label{alg:entry2exit:in-scope-end}
     \Else
        \State $\proj \gets \project(\vars(\literal), \exit)$ \label{alg:entry2exit:builtin-beg}
        \State $\txt{MaybeAbs} \gets \builtin(\literal, \proj)$ \label{alg:entry2exit:builtin}
        \If{$\txt{MaybeAbs} = \txt{fail}$} \label{alg:entry2exit:builtin-not-know}
          \State $\asub{\txt{exit}} \gets \topmost(\vars(\literal), \proj)$\label{alg:entry2exit:builtin-topmost}
        \Else
          \State $\asub{\txt{exit}} \gets \txt{MaybeAbs}$ 
        \EndIf
          \State $\exit \gets \extend(\exit, \asub{\txt{exit}})$ \label{alg:entry2exit:builtin-end}   
    \EndIf
    \EndFor
    \State \Return $\exit$ \label{alg:ana-entry2exit-end}
\EndFunction
\end{algorithmic}
\vspace*{-0.5mm}
\end{algorithm}

\vspace*{-5mm}
\section{
  Environment Reassociation and 
  Abstract Environment Trimming
}
\label{sec:reassociation-and-trimming}
\vspace*{-2mm}
Given a clause $H\pneck B_1,\ldots, B_n$, its \emph{environment} is the
set of all the variables appearing in the clause, defined as
$\vars(H)\cup \bigcup_{i=1}^{n} \vars(B_i)$.  
As mentioned in
Section~\ref{sec:sharing}, 
a
set-sharing abstraction at any analysis point contains, at most, $2^V
- 1$ sharing sets, where $V$ is the size of the environment.
Intuitively, a clause
should be faster to analyze if it has fewer variables.
Since it is not possible to artificially reduce the number of
variables present in a clause, we propose two techniques:
rearranging the literals of the body into new predicates,
and
modifying the domain of the abstraction without altering the clause
being analyzed.
\vspace*{-4mm}\subsection{Environment Reassociation}\vspace*{-2mm}
\emph{Expression reassociation} \citep{Briggs1994EffectivePR}, also known as
reordering or restructuring, is a technique 
that
involves changing the grouping of terms in an expression without
altering its overall value. It is
used for optimization
purposes, such as improving performance, reducing floating-point
errors, eliminating redundancies, etc. 
\\
Given a clause $H\pneck B_1,\ldots, B_n$, 
a \emph{partition} 
is a collection $P_1,\ldots,P_s$ such
that each $P_j$ is a
subsequence of \emph{consecutive} literals, $P_j=B_{i}, B_{i+1},\ldots,B_{k}$
with $1\leq i<k \leq n$, such that given $P_j$,$P_{j+1}$ with
$j\in\{1,s-1\}$: a) if the first element of $P_{j+1}$ is $B_k$, then
the last element of $P_{j}$ is $B_{k-1}$, b) $B_1 \in P_1$ and c)
$B_n \in P_s$.

\emph{Folding} each of the literals encapsulated by each $P_i$
generates new auxiliary predicates whose environments are smaller (or
equal) than the environment of the original predicate. %
This
procedure can be repeated recursively over the auxiliary predicates
obtaining a new collection of predicates with reduced environments.
Finally, our focus is to find an
\emph{optimal} partition. An optimal partition is a (possibly
recursive) partition where the number of variables in the
environments of each of the auxiliary predicates generated is minimal.

Consider the clause of predicate \txt{qplan/3} shown in
Figure~\ref{code:qplan} (the meaning of the comments will be explained later).
We refer to body 
literals by $\txt{L}_i$, where $i$ is a position in the body of the clause.
For example,
$\txt{L}_4$=$\txt{mark(P0,L,0,V1)}$. \\
The collections $P_1$=$\txt{L}_1,\ldots, \txt{L}_3$, $P_2$=$\txt{L}_4,\ldots,
\txt{L}_6$, and $P_3$=$\txt{L}_7,\ldots, \txt{L}_9$ define a
partition of the 
predicate \txt{qplan}. This partition, when folded,
generates three auxiliary predicates:
$\txt{aux1}(\txt{P0,X0,Vg,N}) \pneck \txt{L}_1,\ldots, \txt{L}_3$,
$\txt{aux2}(\txt{P0,Vg,P2}) \pneck \txt{L}_4,\ldots, \txt{L}_6$, and
$\txt{aux3}(\txt{N,X0,P2,X,P}) \pneck \txt{L}_7,\ldots, \txt{L}_9$,
with environments 
containing 5, 6, and 6 variables respectively. Finally, Figure~\ref{fig:transf-qplan} presents a transformation of \txt{qplan}
obtained by
recursively reassociating the predicate.
Each auxiliary
clause is annotated with the worst case size for any
set-sharing abstraction traversing it.
While in the original program the maximum size is $2^{12}$-$1$, it is
reduced to $2^6$-$1$ in the transformed program.
\captionsetup[sub]{labelfont=bf,labelformat=Anumber, labelsep=space}
\begin{figure}
  \caption{\txt{qplan/3} predicate and its environment trimming-based
    transformation.}\vspace{-3mm}
\centering
\begin{subfigure}[b]{\textwidth}
  \caption{
    \txt{qplan/3}
definition (with annotations on live variables).\label{code:qplan}}
\prettylstciao
\begin{lstlisting}[mathescape=true] 
qplan(X0,P0,X,P) :-          % $\textcolor{lightred}{\# \asub{} < 2^{12}}$
   numbervars(X0,0,I),       % I alive
   variables(X0,0,Vg),       % I,Vg alive
   numbervars(P0,I,N),       % I,Vg,N alive
   mark(P0,L,0,Vl),          % Vg,N,L,Vl alive and I dies 
   schedule(L,Vg,P1),        % Vg,N,L,Vl,P1 alive
   quantificate(Vl,0,P1,P2), % N,Vl,P1,P2 alive and Vg,L die
   functor(VA,f,N),          % N,P2,VA alive and Vl,P1 die
   melt(X0,VA,X),            % P2,VA alive and N dies
   melt(P2,VA,P).            % P2,VA alive
 \end{lstlisting}
\vspace*{-2mm}
\end{subfigure}
\begin{subfigure}[b]{\textwidth}
\begin{minipage}{0.47\linewidth}
\prettylstciao
\begin{lstlisting}[mathescape=true]
qplan(X0,P0,X,P) :- % $\textcolor{lightred}{\# \asub{} < 2^6}$
    qplan$_{\mtit{aux}1}$(X0,P0,P2,VA),  
    qplan$_{\mtit{aux}2}$(X0,X,P,P2,VA).

qplan$_{\mtit{aux}1}$(X0,P0,P2,VA) :- % $\textcolor{lightred}{\# \asub{} <2^5}$
    qplan$_{\mtit{aux}11}$(X0,P0,P2,N),  
    functor(VA,f,N).          

qplan$_{\mtit{aux}11}$(X0,P0,P2,N) :-  % $\textcolor{lightred}{\# \asub{} <2^5}$
    qplan$_{\mtit{aux}111}$(X0,P0,N,Vg),
    qplan$_{\mtit{aux}112}$(P0,P2,Vg).   

qplan$_{\mtit{aux}111}$(X0,P0,N,Vg) :-  % $\textcolor{lightred}{\# \asub{} <2^5}$
    qplan$_{\mtit{aux}1111}$(X0,Vg,I),
    numbervars(P0,I,N).     
\end{lstlisting}
\end{minipage}
\hspace*{6mm}
\begin{minipage}{0.47\linewidth}
\prettylstciao
\begin{lstlisting}[mathescape=true,firstnumber=16]
 qplan$_{\mtit{aux}1111}$(X0,Vg,I) :- % $\textcolor{lightred}{\# \asub{} <2^3}$
    numbervars(X0,0,I),  
    variables(X0,0,Vg).  

qplan$_{\mtit{aux}112}$(P0,P2,Vg) :- % $\textcolor{lightred}{\# \asub{} <2^5}$ 
    mark(P0,L,0,Vl),          
    qplan$_{\mtit{aux}1121}$(P2,Vg,L,Vl).

qplan$_{\mtit{aux}1121}$(P2,Vg,L,Vl) :- % $\textcolor{lightred}{\# \asub{} <2^5}$
    schedule(L,Vg,P1),         
    quantificate(Vl,0,P1,P2).
    
qplan$_{\mtit{aux}2}$(X0,X,P,P2,VA) :- % $\textcolor{lightred}{\# \asub{} <2^5}$
    melt(X0,VA,X),    
    melt(P2,VA,P).  
\end{lstlisting}
\end{minipage}
\vspace*{-2.5mm}
\caption{Optimal transformation of \txt{qplan/3} with the minimal number of non-existential variables in
environments. The maximum size of the set-sharing abstractions is
included as a comment.
  \label{fig:transf-qplan}}
\vspace*{-2mm}
\end{subfigure}
\end{figure}
\vspace{-3mm}
\vspace*{-2mm}
\subsection{Abstract Environment Trimming}
\vspace*{-2mm}
The technique of environment reassociation described before, allows 
obtaining, given a clause, a collection of clauses where the number of
variables appearing in each one is reduced with respect to the
original one.
However, applying transformations over the program
under analysis may not always be desirable.
In this section we provide an alternative approach,
where the domain of abstractions 
is dynamically modified as the
analysis of a clause processes each body literal.
Although the resulting abstraction domains might not be optimal, this
technique avoids the transformations, since such dynamic domain
modifications are performed as part of the abstract operations.
Given a clause \mbox{$\head\pneck\bd{1},\ldots,\bd{n}$} a variable
\txt{X} is a \emph{live variable} 
while analyzing $\bd{i}$ (that we 
refer to as the analysis point $\bd{i}$) if
$\txt{X} \in \vars(\head)$ 
or there exists $j, k$ $1 \leq j\leq i \leq k \leq n$
such that \txt{X} belongs to both  \vars($\bd{j}$) and \vars($\bd{k}$).
Conversely, a variable $\txt{X}$ is a \emph{dead variable} at analysis
point $\bd{i}$ if it does not appear in the body after such point,
i.e., $\txt{X} \not\in \bigcup_{j=i}^n \vars(\bd{j})$.  Our definition
of liveness
is similar to imperative programming, with the difference
that variables are not reassigned and that variables become live on
the first appearance, since logic variables do not need to be declared
in the body of a clause~\citep[p. 608-610]{dragon-book-2006}.
Figure~\ref{code:qplan} shows, at each program point,
which
body variable lives or dies. 
In that sense it is reminiscent of the concept of \emph{environment
  trimming} used in the Warren Abstract
Machine~\citep{warren-sri-short,hassan-wamtutorial}, but more general,
since variables also come in instead of only out, and the objective is
of course different: optimizing abstract operations vs.\ saving stack
space.
\begin{algorithm}[b ]
  \caption{Functions to detect live and dead variables. \label{alg:live-die}}
\vspace{-0.5mm}
  \begin{algorithmic}[1]
    \Function{live-vars}{$\txt{LiveVars}, \bd{}$}
    \State $\txt{LitVars} \gets \vars(\bd{})$
    \State \Return $\{X \in \txt{LitVars} \; s.t. \; X \notin \txt{LiveVars}\}$
    \EndFunction
    
    \Function{dead-vars}{$\txt{LiveVars}, \txt{HVars}, \{\bd{i},\ldots,\bd{n}\}$}
    \State $\txt{FutVars} \gets \bigcup_{j=1}^n\vars(\bd{j})$
    \State \Return $\{X \in \txt{LiveVars} \; s.t.\;  X \notin \txt{FutVars}\wedge X\notin \txt{HVars}\}$
    \EndFunction
  \end{algorithmic}
\vspace*{-1mm}
\end{algorithm}%
\begin{algorithm}
  \caption{
    Version of \entrytoexit~dynamically modifying the abstraction domain.}
  \label{alg:entry_to_exit_trim}
  \begin{algorithmic}[1]
    \Function{\entrytoexit}{$\entry, \head, \{\bd{1},\ldots,\bd{m}\}$}
     \State $\txt{HVars} \gets \vars(\head)$ \hfill\Comment{Obtain the Head Variables}
     \State $\txt{LiveVars} \gets \txt{HVars}$ \hfill\Comment{Initialize the Live Variables}\label{alg:init-livs}
     \State $\exit \gets \entry$
    \For{$i \in \{1,\ldots,m\}$}
    \State $\txt{NLiveVars} \gets \Call{live-vars}{\dom(\entry), \bd{i}}$ \hfill\Comment{Get the New Live Vars.} \label{alg:bec-alive}
    \State $\txt{LiveVars} \gets \txt{LiveVars}\cup\txt{NLiveVars}$\hfill\Comment{Update the Live Variables} \label{alg:upd-live1}
    \State $\entry^{\txt{aug}} \gets \augment(\txt{NLiveVars}, \entry)$\hfill\Comment{Augment the Abstraction} \label{alg:aug-live}

    \If{$\Call{recursive-call}{\txt{Literal}, \head}$} 
        \State $\exit \gets \Call{compute-fixpoint}{\exit, \bd{i}}$
    \ElsIf{$\Call{predicate-in-scope}{\txt{Literal}}$} 
        \State $\exit \gets \Call{analyze-pred}{\exit, \bd{i}}$ 
     \Else
        \State $\proj \gets \project(\vars(\bd{i}), \exit)$ 
        \State $\txt{MaybeAbs} \gets \builtin(\bd{i}, \proj)$
        \If{$\txt{MaybeAbs} \neq \txt{fail}$}
          \State $\asub{\txt{exit}} \gets \topmost(\vars(\bd{i}), \proj)$
        \Else
          \State $\asub{\txt{exit}} \gets \txt{MaybeAbs}$
        \EndIf
        \State $\exit \gets \extend(\exit, \asub{\txt{exit}})$ 
        \State $\txt{DeadVars} \gets \Call{dead-vars}{\txt{LiveVars}, \txt{HVars},\{\bd{j}\}_{j=i+1}^{m}}$ \hfill\Comment{Get Dead Vars.} \label{alg:dying-vars}        
          \State $\txt{LiveVars} \gets \txt{LiveVars} \backslash \txt{DeadVars}$ \hfill\Comment{Update Live Variables} \label{alg:upd-live2}
          \State $\exit \gets\project(\txt{LiveVars}, \exit)$ \hfill\Comment{Restrict to the Live Variables} \label{alg:proj-lives}
    \EndIf
    \EndFor
    \State \Return $\exit$ 
\EndFunction
\end{algorithmic}
\vspace{-0.5mm}

\end{algorithm}%

Algorithm~\ref{alg:live-die} presents the operations
$\Call{live-vars}{}$ and $\Call{dead-vars}{}$
to obtain the
variables becoming alive and the ones which are dead at a given
analysis point.
$\Call{live-vars}{}$ receives the set of current 
live variables
(\txt{LiveVars}), and the literal that is going to be analyzed
($\bd{}$), and returns the set of
new live variables
at that analysis point, i.e., the variables that appear in the literal
but were not in \txt{LiveVars}.
The other operation, $\Call{dead-vars}{}$,
takes as input the set of current 
live 
variables (\txt{LiveVars}), the variables of
the head (\txt{HVars}),
and the set of literals that have not been analyzed yet
($\{\bd{i},\ldots,\bd{n}\}$), and produces as output
a set containing the variables of \txt{LiveVars} that do not appear in
any of the literals nor belong to the clause head.
More efficient procedures to determine the liveness of variables are
possible. However, we checked experimentally that 
they do not offer substantial improvements, and thus
we decided to keep these simpler definitions.
With these auxiliary operations, the analyzer can
determine, at each analysis point, whether a variable is live or dead.
With this information, it is possible to restrict the domain of the
abstractions to the set of live variables.
To do so, the computation of the \succ~abstraction 
is modified slightly:

\ \\ [-10mm]
\begin{align*}
  \succ  &= \extend(\call,\goal,\prime) \\[-1.2mm]
  \prime &= \exittoprime(\project(\vars(\head), \exit), \head, \goal) \\[-1.2mm]
  \exit  &= \entrytoexit(\entry, \head, \body) \\[-1.2mm]
  \entry &= \calltoentry(\proj, \goal, \head) \\[-1.2mm]
  \proj  &= \project(\vars(\goal), \call).
\end{align*}
In this case, 
the \entry~abstraction is obtained
by 
directly computing \calltoentry,
instead of by augmenting the result of the \calltoentry~invocation, 
as was done in the schema presented in Section~\ref{sec:plai}. Thus,
in this approach, the domain of the \entry~abstraction is exactly the
set of head variables (which are the only variables alive at that
analysis point).  Finally, the function \entrytoexit~presented in
Algorithm~\ref{alg:entry_to_exit}~is modified so that it keeps the
abstraction defined only over the live variables.
Such a modified version is described by
Algorithm~\ref{alg:entry_to_exit_trim}.
There, a set containing the live
variables is carried around while analyzing a clause body
($\{\bd{1},\ldots,\bd{m}\}$). Such set is initialized with the
variables of the clause head (Line~\ref{alg:init-livs}), and 
updated by adding new variables to it as they
become alive 
(Lines~\ref{alg:bec-alive}-\ref{alg:upd-live1}) and by removing
the dead variables 
(Lines~\ref{alg:dying-vars}-\ref{alg:upd-live2}).
Accordingly, the abstraction is augmented with the new live variables
(Line~\ref{alg:aug-live}) and reduced when some of them die
(Line~\ref{alg:proj-lives}).
\vspace*{-6mm}
\section{Experimental Evaluation\label{sec:evaluation}}
\label{sec:experiment}
\vspace*{-2mm}
We have conducted an extensive experimental study to assess the
benefits of the techniques proposed, to which we will refer to
here as \emph{reassociation} and \emph{trimming} for short.
In particular, we measured, for a variety of
programs, 
the effects of applying both techniques to
a number of abstract domains that use set sharing:
the classical set-sharing domain,
\txt{share}~\citep{abs-int-naclp89-short}, the combination of sharing
and freeness, \txt{shfr}~\citep{freeness-iclp91-medium}, and the 
combination of sharing and freeness including cliques,
\txt{shfr-clique}~\citep{shcliques-padl06-short}; 
the latter is the most efficient of the three, while \txt{share} and
\txt{shfr} are, in general, more precise but slower. %
All experiments were performed on 
Debian/GNU Linux 12 (bookworm) 64bit (amd64) with 128GB RAM,
and 800 GB of disk space.
We focus on analysis
times since precision is unchanged.

We start by showing in Table~\ref{tab:lpdoc} the detailed results for
one of our benchmarks, the \lpdoc documenter
\citep{lpdoc-reference-manual-3.0-short,lpdoc-cl2000-short}, which is
a relatively large, real-world application, and whose results are typical. The
\lpdoc source code is composed of 26 modules that exhibit different  
challenges: for some of them analysis terminates in times that range from a few
milliseconds to several minutes, while others cannot be analyzed,
either because of a timeout, set for these experiments at 5 minutes per module, or
by running out of memory. In either of these cases we will say that the
analysis fails.
For each abstract domain, Column \textbf{TC} shows the \emph{total}
time that the classical domain requires to analyze the modules,
columns \textbf{TR}
and
\textbf{TT}
show the \emph{total} analysis times when applying reassociation and
trimming respectively to the corresponding classical domain, and
columns $\rho$\textbf{R} and $\rho$\textbf{T}
present the relative speed-up computed as \textbf{TC}/\textbf{TT} and
\textbf{TC}/\textbf{TR}.
\emph{Total} analysis times are the addition
of the times for the 
loading
of the module,
the pre-processing
(including the transformation required by the reassociation method),
and the actual analysis time.
Some statistics are included at the bottom of the table: the total
number of modules (\textbf{Mods}) and the number of modules for which
the different analyses fail, for the classical approach (\textbf{FC}),
when applying reassociation (\textbf{FR}), and when applying trimming
(\textbf{FT}).
Finally, $\mu$\textbf{T} and $\mu$\textbf{R} show the mean of the 
speed-ups obtained.
We can see that
applying reassociation and trimming allows analyzing a
significant number of modules that could not be analyzed with the
classical techniques. In particular, when applied to the 
\txt{shfr-clique} domain they achieve a full analysis of the \lpdoc
source code.
The mean speed-ups show that trimming outperforms reassociation in
this application. This is due to the significant improvements that
trimming brings to the modules \textbf{images} and \textbf{refsdb}.
Specifically, in the case of \textbf{images}, the trimming approach
significantly reduces analysis times for the \txt{share} and
\txt{shfr} domains, resulting in speed-ups comparable to those
achieved by \txt{shfr-clique}. However, trimming introduces a
slow-down when applied to the \txt{shfr-clique} domain. Conversely,
reassociation does not provide any significant changes.
In the \textbf{texinfo} module, reassociation performs 10 times
faster than trimming, while in the remaining modules, both methods
behave similarly. Our hypothesis is that the overhead introduced by
the transformation is what causes reassociation to slightly
under-perform in some cases.

The proposed methods were also evaluated across a set of classic
benchmarks (see Table~\ref{tab:benchs} in the appendices).
The benchmarks include a variety of examples, ranging from simple
predicates that feature only direct recursions, such as \textbf{qsort}
and \textbf{append}, to more complex cases with mutual recursions and
elaborate aliasing. Some benchmarks are extracted from real-world
programs. For example, \textbf{aikal} is part of an analyzer for the
AKL language, while \textbf{read} and \textbf{rdtok} are Prolog
parsers. Additionally, parts of actual programs are also included,
such as \textbf{ann} (the \&-Prolog parallelizer), \textbf{qplan} (the core of
the Chat-80 application), and \textbf{witt} (a conceptual clustering
application).
As expected, in these comparatively less challenging programs the
advantages obtained are smaller, but they are still significant. For
instance, using trimming and reassociation 
bring mean relative speed-ups of $1.1$ and $1.06$ respectively
for 
the \txt{share} domain; $0.95$ and
$1.5$ 
for 
\txt{shfr}; and $2.2$ and $1.97$
for 
\txt{shfr-clique}.
\begin{landscape}
\begin{table}[htbp]
\begin{small}
  \vspace{-1mm}
  \caption{Analysis times and statistics for the source code of \lpdoc. Times are in mS.\label{tab:lpdoc}}
  \vspace*{-2mm}
  \begin{tabular}
    {||p{18mm}||p{11mm}|p{11mm}|p{9mm}|p{11mm}|p{9mm}||p{11mm}|p{11mm}|p{9mm}|p{11mm}|p{9mm}||p{11mm}|p{11mm}|p{9mm}|p{11mm}|p{9mm}||}
 \hline\hline
 & \multicolumn{5}{c||}{\textbf{share}} & \multicolumn{5}{c||}{\textbf{shfr}} & \multicolumn{5}{c||}{\textbf{shfr-clique}} \\ \hline
 \textbf{module}  & \textbf{TC}      & \textbf{TR} & \textbf{$\rho$R}  & \textbf{TT} & \textbf{$\rho$T}   & \textbf{TC}      & \textbf{TR} & \textbf{$\rho$R}   & \textbf{TT} & \textbf{$\rho$T}   & \textbf{TC}     & \textbf{TR} & \textbf{$\rho$R} & \textbf{TT} & \textbf{$\rho$T} \\ \hline
 \footnotesize
html         &   23485 &    1377 & 17.04 &    2646 &   8.87 &   20858 &    1373 & 15.19 &    2516 &   8.29 &    1522 &   1293 &  1.18 &   1532 &  0.99 \\
html\_assets &    6.20 &    8.20 &  0.76 &    6.75 &   0.92 &    7.25 &    6.46 &  1.12 &    7.79 &   0.93 &   13.70 &  12.42 &  1.10 &  14.25 &  0.96  \\
aux          &    4.12 &    4.60 &  0.90 &    4.09 &   1.01 &    4.51 &    5.17 &  0.87 &    4.58 &   0.98 &    5.57 &   6.75 &  0.83 &   5.82 &  0.96  \\
man          & timeout &   32944 &     - &   64490 &      - & timeout &   29235 &     - &   60521 &      - &   23.74 &  27.23 &  0.87 &  23.32 &  1.02  \\
doctree      &   10617 &    7095 &  1.50 &    5807 &   1.83 &    6426 &    5443 &  1.18 &    1898 &   3.39 &    4485 & 555.77 &  8.07 & 568.98 &  7.88  \\
docmod       &    0.82 &    0.76 &  1.07 &    0.85 &   0.97 &    0.75 &    0.88 &  0.85 &    0.77 &   0.97 &    0.75 &   0.97 &  0.87 &   0.70 &  1.08  \\
images       &  132364 &  134906 &  0.98 &  134.10 & 987.03 &  136011 &  130448 &  1.04 &  123.82 &   1098 &   16.95 &  14.83 &  1.14 &  52.66 &  0.32  \\
messages     &    3.75 &    3.58 &  1.05 &    4.29 &   0.87 &    4.22 &    4.39 &  0.96 &    4.32 &   0.98 &    6.08 &   5.47 &  1.06 &   5.96 &  1.02  \\
structure    &   15.59 &   12.63 &  1.23 &   11.54 &   1.35 &   17.77 &   14.07 &  1.26 &   13.27 &   1.34 &   21.44 &  16.34 &  1.31 &  18.24 &  1.18  \\
lpdoc\_sing. &  342.13 &   17.53 & 19.51 &   12.58 &  27.19 &  331.32 &   19.85 & 16.96 &   14.21 &  23.31 &  533.04 &  29.56 & 18.03 &  22.26 & 23.94  \\
docmaker     &   40.49 &   34.69 &  1.17 &   33.97 &   1.19 &   41.71 &   46.96 &  0.89 &   33.44 &   1.25 &   60.92 &  76.91 &  0.79 &  69.32 &  0.88 \\
bibrefs      &  119737 &   64348 &  1.86 &   91029 &   1.32 &   74779 &   12024 &  6.22 &   40949 &   1.83 &    1560 & 590.09 &  2.64 & 780.31 &  2.00  \\
html\_templ. &  124.43 &   34.46 &  3.61 &   26.27 &   4.74 &  121.22 &   25.99 &  4.66 &   22.03 &   5.50 &   41.23 &  42.65 &  0.97 &  46.61 &  0.88  \\
lpdoc\_help  &    3.01 &    3.56 &  0.85 &    2.73 &   1.10 &    3.84 &    4.22 &  0.91 &    3.01 &   1.28 &    5.34 &   5.82 &  0.92 &   4.43 &  1.21  \\
texinfo      & timeout & timeout &     - & timeout &      - & timeout & timeout &     - & timeout &      - & timeout & 171.50 &     - &   1440 &     -  \\
comments     &   35.68 &   38.58 &  0.93 &   22.50 &   1.59 &   25.90 &   24.91 &  1.04 &   23.46 &   1.10 &   44.29 &  49.59 &  0.89 &  36.56 &  1.21  \\
errors       &    0.79 &    0.95 &  0.83 &    0.74 &   1.07 &    0.97 &    0.76 &  1.29 &    0.76 &   1.29 &    0.00 &   1.06 &  0.09 &   1.14 &  0.09  \\
parse        &   55513 &   54284 &  1.02 &   18022 &   3.08 &   49511 &   49577 &  1.00 &   15353 &   3.22 &    2381 &   2404 &  0.99 &   2492 &  0.96  \\
autodoc      & timeout & timeout &     - & timeout &      - & timeout & timeout &     - & timeout &      - & timeout &   6242 &     - &   6864 &     -  \\
index        &    1856 &  395.24 &  4.70 &  685.04 &   2.71 &    1298 &  344.32 &  3.77 &  562.93 &   2.31 &   93.31 &  66.97 &  1.39 &  95.55 &  0.98  \\
filesystem   &   80.38 &   41.11 &  1.96 &   53.51 &   1.50 &   80.32 &   51.71 &  1.55 &   51.93 &   1.55 &   59.12 &  65.75 &  0.90 &  63.98 &  0.92  \\
doccfg       &    3.06 &    3.53 &  0.87 &    2.18 &   1.40 &    4.03 &    3.75 &  1.08 &    2.49 &   1.62 &    3.03 &   3.54 &  0.85 &   0.44 &  6.82  \\
refsdb       & timeout &    1507 &     - &    2991 &      - &   97871 &    1093 & 89.54 &  572.88 & 170.84 &   26853 &   1433 & 18.73 & 393.76 & 68.20  \\
lookup       &   11.68 &   12.04 &  0.97 &    6.98 &   1.67 &   14.44 &   12.54 &  1.15 &    4.01 &   3.60 &   25.30 &  18.14 &  1.39 &  13.09 &  1.93  \\
version      &    0.28 &    0.23 &  1.23 &    0.30 &   0.92 &    0.32 &    0.30 &  1.05 &    0.31 &   1.04 &    0.28 &   0.32 &  0.88 &   0.30 &  0.93  \\
settings     &   15.36 &   17.97 &  0.85 &   17.19 &   0.89 &   19.17 &   22.26 &  0.86 &   20.03 &   0.96 &   37.77 &  42.99 &  0.88 &  41.01 &  0.92  \\
nil          &    2.11 &    2.60 &  0.81 &    2.27 &   0.93 &    3.52 &    2.26 &  1.56 &    2.43 &   1.45 &    1.56 &   1.63 &  0.96 &   1.69 &  0.92  \\
state        &   38271 &    4745 &  8.06 &    4782 &   8.00 &   38006 &    4647 &  8.25 &    4733 &   8.03 &    1177 &   1103 &  1.07 &   1103 &  1.07  \\
\hline \hline
& \multicolumn{5}{c||}{\textbf{share}} & \multicolumn{5}{c||}{\textbf{shfr}} & \multicolumn{5}{c||}{\textbf{shfr-clique}} \\ \hline
\footnotesize
\textbf{Mods} & \multicolumn{5}{c||}{27} & \multicolumn{5}{c||}{27}  & \multicolumn{5}{c||}{27}\\ 
\textbf{FC} & \multicolumn{5}{c||}{4} &  \multicolumn{5}{c||}{5}& \multicolumn{5}{c||}{2} \\ 
\textbf{FR} &  \multicolumn{5}{c||}{2}& \multicolumn{5}{c||}{2} & \multicolumn{5}{c||}{0} \\ 
\textbf{FT} & \multicolumn{5}{c||}{2} & \multicolumn{5}{c||}{2} & \multicolumn{5}{c||}{0} \\ 
$\mu$\textbf{R} & \multicolumn{5}{c||}{3.07}& \multicolumn{5}{c||}{8.13} & \multicolumn{5}{c||}{2.64} \\
$\mu$\textbf{T} & \multicolumn{5}{c||}{44.26} & \multicolumn{5}{c||}{53.82} & \multicolumn{5}{c||}{4.97} \\ \hline
  \end{tabular}
\end{small}

 \end{table}
\end{landscape}
Most of these modules already required small analysis times 
(less than $0.1$ seconds, which
may indicate they offer fewer opportunities for optimization).
The improvements are most significant for \txt{qplan}, which had
the largest analysis times to begin with:
$5.7$ seconds (\txt{share}), $1$ second (\txt{shfr}),
and $5$ seconds (\txt{shfr-clique}).
In this case, trimming and
reassociation bring analysis times of $1.1$ and $0.5$ seconds,
respectively, with \txt{share}; and $0.3$ and $0.2$ seconds with
\txt{shfr}.
Moreover, with \txt{shfr-clique} both techniques bring the time down
to $0.1$ seconds ($50\times$ speed-up, the largest one).
The
full experimental results can be found in Appendix~\ref{app:class}.

Given the positive results, in a second phase of our experimentation
we decided to greatly expand the code base used to measure the impact
of applying abstract environment trimming, in order to explore whether
the improvements obtained for the classic benchmarks and the \lpdoc
application would translate much more widely.
All in all we have analyzed around 1200 program modules.
These include, in addition to the previously mentioned 
\lpdoc documenter (28 modules) and classic benchmarks (26
modules): all the libraries distributed with the \ciao
system~\citep{hermenegildo11:ciao-design-tplp-medium}
(comprising more than 1000
modules),   
\ciaopp's
program analyzer~\citep{ciaopp-sas03} \plai (56 modules),
s(CASP)~\citep{scasp-iclp2018}, a top-down interpreter for
ASP programs with constraints (44 modules), and 
Spectector~\citep{spectector}, a tool for automatically detecting
leaks caused by speculatively executed instructions in x64 assembly
programs (15 modules).
The $\approx$1000 library modules in the standard \ciao distribution come
from many sources including libraries ported from
SICStus~\citep{sicstus-journal-2012}; libraries common with
Yap~\citep{yap-journal-2012-short},
XSB~\citep{xsb-journal-2012-short},
SWI~\citep{swi-journal-2012-short}, etc.,
including those from the
Prolog Commons project ({\small \url{https://prolog-commons.org/}}); 
the classic libraries from O'Keefe and those for Constraint Logic
Programming over Rationals and Reals~\citep{holzbaur-clpqr};
libraries for implementing language extensions such as functional
syntax or the \ciao assertion language; etc.,
developed by around 100 different programmers.
We argue that this selection of benchmarks
constitutes a good representation of
real-life code written in Prolog.
The detailed results of these experiments can be found
in~\ref{app:tables}, while here we present them in a more 
manageable aggregated form.

\begin{figure}
  \includegraphics[width=0.9\textwidth]{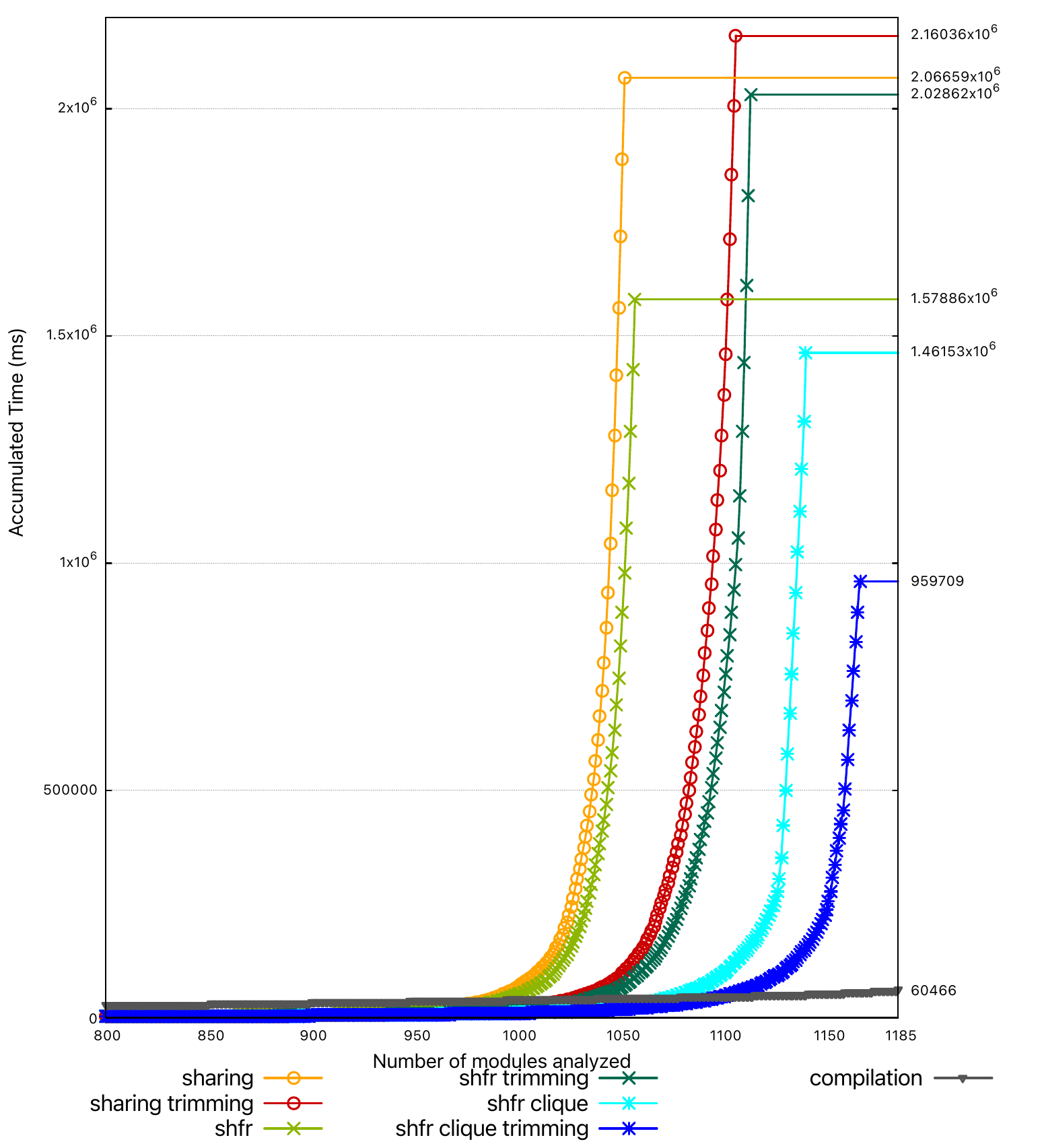}
  \vspace*{-4mm}
  \caption{Cactus plot aggregating all the benchmarks
    (1185 modules).
    Times are in mS. \label{figure:cactus-plot}}
  \vspace*{-4mm}
\end{figure}
We first present in Figure~\ref{figure:cactus-plot} a \emph{cactus plot}
of the results.
Cactus plots display,
on the $X$-axis,
the number of benchmarks that can be analyzed, i.e., those for which the
analysis does not time out or run out of memory, and, on the $Y$-axis,
the \emph{accumulated} analysis time. The plots for each abstract domain and
analysis technique are generated by taking all the analysis times,
\emph{sorting them in increasing order}, and then plotting them following the
formula $(i, t_i)$ 
where $t_i$ is the \emph{accumulated} time in the
$i$-th position. This way, the analysis time of the benchmark
corresponding to the $i+1$-th position is $t_{i+1} - t_i$. Darker
colors represent abstract domains with trimming, while lighter colors
represent domains with classic analyses. Empty circles correspond to
the classic set-sharing domain, crosses to \txt{shfr}, and stars to
\txt{shfr-clique}. The module compilation times are given for comparison, represented by gray
triangles --resulting in the gray vertical line. The plot has been zoomed in to exclude points with
$X$-values (numbers of benchmarks)
less than $800$, as the most interesting information is in the more
challenging benchmarks beyond, that require the highest
analysis times. 
Also, plots corresponding to modules for which the
analysis fails are excluded. This figure allows us to observe that when
the domains are used with abstract trimming, they perform better than
their classical counterparts. Specifically, \txt{share} and \txt{shfr}
reduce the number of modules they are unable to analyze from
$134$ and
$130$ to $81$ and $74$, respectively, corresponding to a $39.55$\% and
$43.07$\% improvement.
For \txt{shfr-clique} applying abstract environment
trimming results in failure to analyze only
$21$ modules versus the
$47$ that the classical approach failed to analyze, while reducing the
accumulated time by $8.36$ minutes. This translates to a reduction in
timeouts by $55.32$\% while reducing the
accumulated time by $34.3$\%.
We have also obtained mean speed-ups of
$5.82$, $5.91$, and $22.08$ when analyzing with \txt{share}, \txt{shfr}
and \txt{shfr-clique} respectively. Moreover, the number of modules
that can be analyzed in a time lower than the compilation time also
increases.

\begin{figure}
  \includegraphics[width=\textwidth]{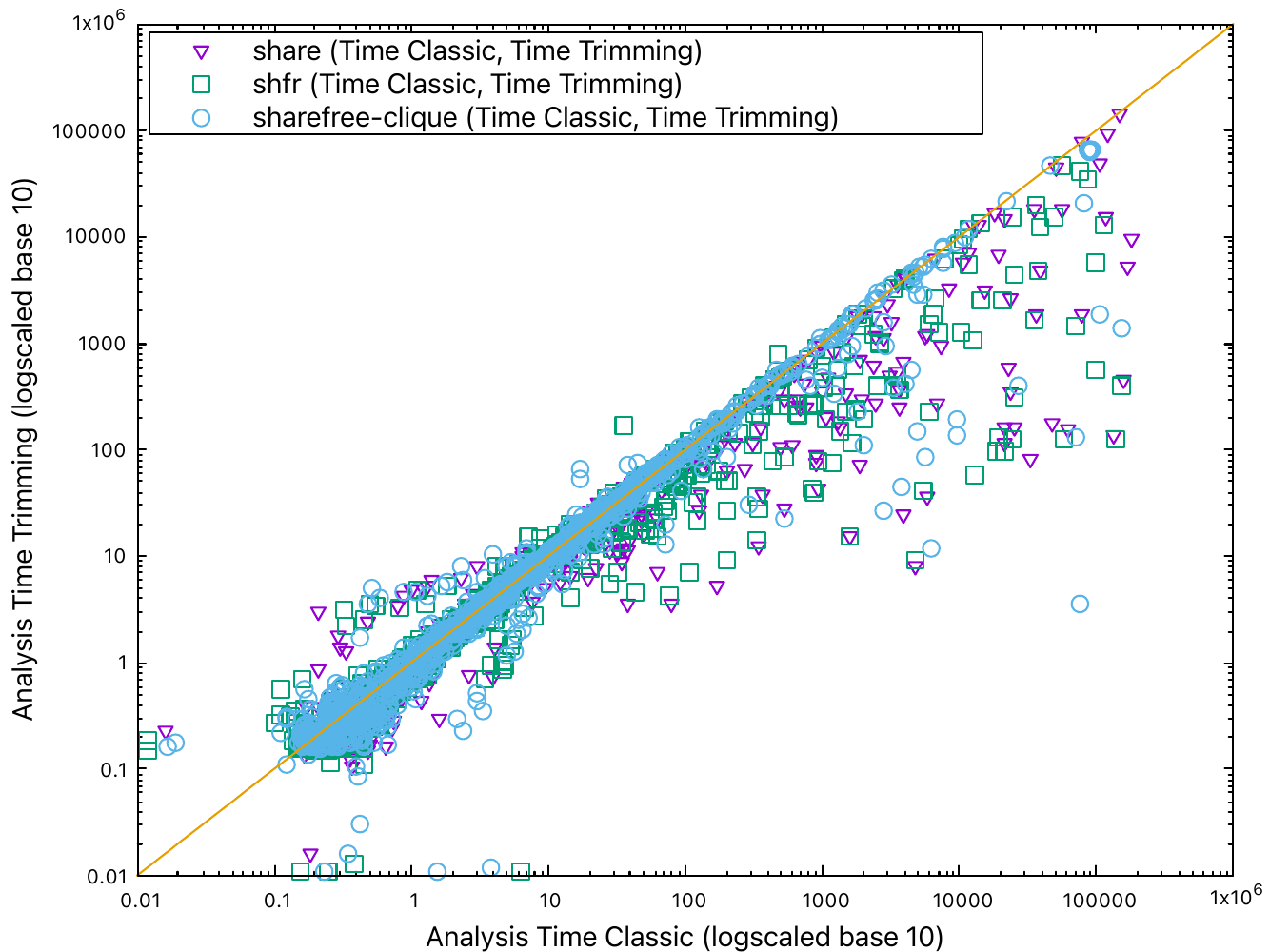}
  \vspace*{-7mm}
    \caption{Scatter plot comparing absolute analysis times (in mS).\label{figure:scatter-plot}}
   \vspace*{-2mm}
\end{figure}

\begin{figure}
  \hspace*{-9mm}
  \includegraphics[width=1.1\textwidth]{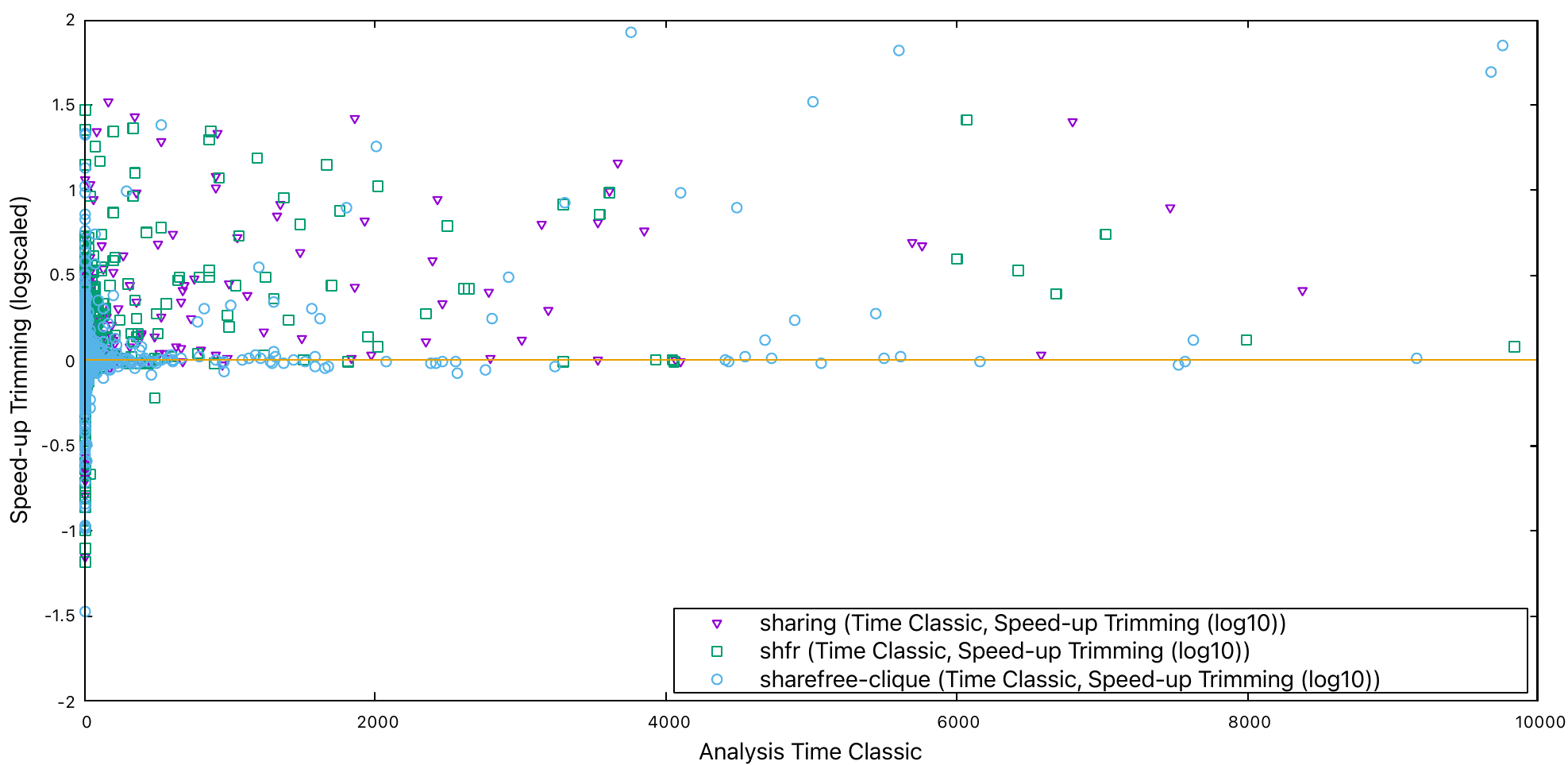}
  \vspace*{-7mm}
  \caption{Scatter plot showing classic analysis time (in mS) vs.
    speed-up obtained (logscaled base 10).\label{figure:scatter-plot-spd}}
   \vspace*{-2mm}
\end{figure}

The cactus plot shows how the analysis results accumulate. In order to
show how these results relate individually,
Figure~\ref{figure:scatter-plot} presents a scatter plot displaying
the time required to analyze each module.
Given a point $(x, y)$, the value in $x$ corresponds to the time
required by the classical analysis to analyze a given module, while
$y$ corresponds to the time required to analyze that same module using
abstract trimming. Modules that are not analyzed with the classic
approach or with both are not displayed. If the points are close to
the orange line, it means that the times are very similar; if they are
above the line, it means that abstract trimming introduces an
overhead; if they are below, abstract trimming speeds up the analysis.
To complement this information,
Figure~\ref{figure:scatter-plot-spd} presents a scatter plot
displaying a comparison between the time required to analyze each
module with the classical analysis ($X$-axis) and the speedup
obtained by applying the abstract trimming technique ($Y$-axis).  The
speedup values range from very close to zero to significantly larger
numbers
(see the 0.09 and the 1098 speedup obtained by abstract trimming
when analyzing the ``errors'' module with sharefree-clique and the
``images'' module with shfr, as shown in Table~\ref{tab:lpdoc}).  To
better represent these values, we have applied a base 10 logarithm
to the resulting speedups. Thus, values between 0 and 1 become
negative (with larger absolute values the closer they are to 0),
while values greater than 1 are scaled in the positive plane. 
The most significant performance improvements are observed in the
right-most side of the figures, corresponding to the
modules where the classical approach takes more time. Conversely, in
cases where the classical approach is very fast (left-most part of the
figures), the technique of abstract trimming does not yield many
speedups but introduces some overheads, which are however small.
Another observation is that in most benchmarks, the analysis times are
quite low, but of course our target has been the rest that present
significant challenges. 
\begin{table}
  \caption{Complete statistics.
    \label{tab:stats:all}}
  \begin{center}
  \vspace*{-5mm}
  \begin{tabular}{||l|r|r|r||}
  \hline \hline
  & \textbf{share} & \textbf{shfr} & \textbf{shfr-clique} \\ \hline  \hline
  \textbf{Modules} & 1186  & 1186  & 1186 \\ 
  \textbf{FC}      & 134   & 130   & 47  \\ 
  \textbf{FT}      & 81    & 74    & 21  \\ 
   $\mu$\textbf{T}  & 5.82  & 5.91 & 22.08  \\ 
   $\mu$\textbf{T} > 1s (70-74-160) & 62.13 & 57.64 & 146.15 \\ 
   $\mu$\textbf{T} > 0.5s (93-97-180) & 47.86 & 45.05 & 130.66 \\ 
   $\mu$\textbf{T} > 0.1s  (143-147-230) & 32.11 & 30.60 & 102.65 \\ 
   $\mu$\textbf{T} < 0.1s (909-913-999) & 1.68 & 2.10 & 24.12 \\ \hline
  \end{tabular}
\end{center}
\vspace*{-3mm}  
\end{table}
To concentrate on this set we have collected the 
speed-up results
considering 
only the benchmarks taking more than 0.1,
0.5, and 1 second when analyzed with \txt{share} (which represents the
slower domain).  These results are presented in
Table~\ref{tab:stats:all}.  For example, the row starting with
\mbox{``$\mu$\textbf{T} > 1s (70-74-160)''}
shows the mean speedup for the benchmarks successfully
analyzed such that the time required to analyze them with \txt{share} is
greater than 1 second or \txt{share} times out. The results show that in
the case of \txt{share},
70 of these more challenging modules are successfully analyzed 
with both trimming and reassociation
(note that both approaches need to succeed in order to compute the
means), 74 in the case of \txt{shfr}, and 160 in the case of
\txt{shfr-clique}.
Similar very positive results are obtained for the other cases. 

\vspace*{-5mm}
\section{Conclusions}
\label{sec:conclusions}
\vspace*{-1mm}

We have proposed a number of techniques for addressing the scalability
problems inherent to set-sharing analyses.
We have focused on the root of the problem: the
potentially exponential dependency of 
the size of the abstractions on the number of variables. We have
cast 
this problem as an instance of expression reassociation and
provided an optimal solution using program
transformations. Additionally, we have proposed a practical solution
that can be integrated into top-down analyzers, based on the liveness
of variables in the body of the clause being analyzed.
We have conducted an extensive experimental evaluation of over 1100
program modules taken from both production code and classical
benchmarks.
We have obtained significant speed-ups, and, more importantly, the
number of modules that require a timeout was cut in half. As a result,
many more programs can be analyzed precisely in reasonable times.
We believe that the results obtained suggest that the
proposed local technique 
improves significantly the scalability of set-sharing analyses, and
can thus enhance 
the practicality of top-down set sharing analysis for production code.
As a possible avenue for future work, note that the
definition of live variables used in this work is local to each
clause of the predicate being analyzed. Future lines of work could
explore
a more global notion that also considers the calls to
predicates
within the clause under analysis. This will presumably incur
additional cost but could also possibly allow further reduction in
the size of the domains of the sharing abstractions.
\vspace*{-6mm}
\small
\bibliographystyle{tlplike}
\bibliography{extracted.bib}
\clearpage{}%
\appendix
\section{An Additional Top-down Sharing Analysis Example}

The example provided in Section~\ref{sec:prelim} is chosen to be
simple in order to make it easy to follow. 
To complete the
discussion about how the analyzer proceeds, we provide a richer
example showing how a complex sharing pattern is computed. This
example will also allow us to show how sharing-set representations
easily grow.
\begin{figure}
  \prettylstciao
  \begin{lstlisting}
app([],L,L).
app([H|T],L0,L2) :-
    app(T,L0,L1),
    L2 = [H|L1].
  \end{lstlisting}
  \vspace*{-4mm}
  \caption{\label{code:app}}
\end{figure}
Let \txt{app/3} be a predicate that succeeds if its third argument is
the list resulting from appending the first two arguments as showed in
Figure~\ref{code:app}.  Let this predicate be called with
\goal\,$=\txt{app([A],[B,C],[A,B,D])}$, and calling abstraction
$\mbox{\call\,=\{\{\txt{A,B}\}, \{\txt{C}\}, \{\txt{D,E}\}\}}$. Thus, the
success abstraction is computed as follows:
\begin{itemize}
\item[i)] \proj=\project($\{\txt{A,B,C,D}\}$, \call)$\,=\{\{\txt{A,B}\}, \{\txt{C}\}, \{\txt{D}\}\}$.
\item[ii)]
  \asub{\txt{entry}}=\calltoentry$(\proj,\goal,\txt{app([H|T],L0,L2)})=\{\{\txt{L0}\},\{\txt{L0,L2}\},\{\txt{L0,L2,H}\},\{\txt{L2}\}\}$
  Notice how, in this case, the head of the second clause has been
  selected instead of the head of the first clause
  (\txt{app([],L,L)}). This is because the analyzer only selects
  the clauses which can unify with the goal.
\item[iii)]
  \entry=\augment$(\{\txt{L1}\},
  \asub{\txt{entry}})=\{\{\txt{L0}\},\{\txt{L0,L2}\},\{\txt{L0,L2,H}\},\{\txt{L1}\},
  \{\txt{L2}\}\}$
\item[iv)] Now the abstracion \exit~is computed by calling
  \Call{entrytoexit}{\entry,\head,\body} as follows:
  \begin{itemize}
  \item[a)] Since the first literal (\txt{app(T,L0,L1)}) is a
    recursive call the fixpoint has to be computed. In a very
    simplified way: both clauses would be analyzed with
    \goal=\txt{app(T,L0,L1)} and
    $\call_1$=\project$(\{\txt{T,L0,L1}\}, \entry)=\{\{\txt{L0}\},\{\txt{L1}\}\}$.
  \item[b)] The analysis of the first clause (\txt{app([],L,L)}),
    creates a sharing between \txt{L0} and \txt{L1} 
    since the unification of \goal=\head~induces 
    \txt{L0=L1}
    . Then, analyzing this clause outputs the prime abstraction
    $\prime_1= \{\{\txt{L0,L1}\}\}$.
  \item[c)] Then, the second clause is analyzed.  The entry is
    computed by computing the abstract unification of
    \txt{app(T,L0,L1)} and \txt{app([H$_r$|T$_r$], L0$_r$, L2$_r$}. We
    have renamed the variables of the clause being analyzed (adding a
    subscript$_r$) to avoid confusion.
    Now, the entry abstraction obtained is
    \entry$_{\txt{fxp}}=\{\{$\txt{L0$_r$}$\},\{$\txt{L2$_r$}$\},
    \{$\txt{L1$_r$}$\}\}$. Notice that, since the abstract unification
    induces the unification \txt{T=[H$_r$|T$_r$]} and \txt{T} is
    ground in the call, both \txt{H$_r$} and \txt{T$_r$} become
    ground.
    In the next invocation to \Call{entrytoexit}{}, the literal
    \txt{app(T$_r$,L0$_r$,L1$_r$)} has to be analyzed with call
    $\{\{$\txt{L0$_r$}$\},\{$\txt{L1$_r$}$\}\}$. This call is the same
    as $\call_1$ (up to variable renaming).
    Thus,
    a fixpoint is reached and the literal is analyzed yielding the
    abstraction $\{\{$\txt{L0$_r$,L1$_r$}$\}\}$. This is extended with
    \entry$_{\txt{fxp}}$ returning:
    $ \{\{$\txt{L0$_r$}$\}, \{$\txt{L2$_r$}$\},\{$\txt{L1$_r$}$\}\}$. Then, the
    next literal (\txt{L2$_r$=[H$_r$|L1$_r$]}) is analyzed and the
    abstraction $\{\{$\txt{L0$_r$},\txt{L1$_r$},\txt{L2$_r$}$\}\}$ obtained.
    Finally, the \exittoprime~invocation returns:
    $\prime_2 = \{\{\txt{L0,L1}\}\}$
  \item[d)] Then, the least upper bound of $\prime_1$ and $\prime_2$ is
    computed. Since they are the same the obtained abstraction is
    $\{\{\txt{L0,L1}\}\}$ as well.
  \item[e)] Such abstraction is used to update the exit abstraction
    carried during the execution of \Call{entrytoexit}{}. Thus, the
    current \exit~abstraction 
    ($\{\{\txt{L0}\},\{\txt{L0,L2}\},\{\txt{L0,L2,H}\},\{\txt{L1}\},
    \{\txt{L2}\}\}$) is updated to
    \exit=$\{\{\txt{L0,L2,H,L1}\},
    \{\txt{L0,L2,L1}\},\{\txt{L0,L1}\},\{\txt{L2}\} \}$ propagating
    the new sharing created between \txt{L0} and \txt{L1}.
  \item[f)] The next literal (\txt{L2=[H|L1]}) is analyzed with
    abstraction \project($\{\txt{L1,H,L2}\}$, \exit)
    =$\{\{\txt{L2,H,L1}\},\{\txt{L2,L1}\},\{\txt{L1}\},\{\txt{L2}\}\}$. This
    literal creates sharing between \txt{L2} and \txt{H} and between
    \txt{L2} and \txt{L1} respectively. Thus, the abstraction obtained is
    $\{\{\txt{L2,H,L1}\},\{\txt{L2,L1}\}\}$ and, after extending, \exit=$\{\{\txt{L0,H,L1,L2}\},\{\txt{L0,L1,L2}\}\}$.
  \end{itemize}
\item[v)] Then, \prime=\exittoprime(\exit, \head,
  \goal)=$\{\{\txt{A,B}\}, \{\txt{A,B,C}\}, \{\txt{A,B,C,D}\},
  \{\txt{A,B,D}\},\{\txt{C,D}\}\}$. Note how, the unification
  \txt{app([H|T],L0,L2)=app([A],[B,C],[A,B,D])} propagates the
  sharings between the variables.
\item[vi)] Finally, the success abstraction is computed by updating
  the call:
  \succ=\extend(\call,\goal,\prime)=$\{\{\txt{A,B}\}, \{\txt{A,B,C}\},
  \{\txt{A,B,C,D,E}\}, \{\txt{A,B,D,E}\},\{\txt{C,D,E}\}\}$.
\end{itemize}
\clearpage
\section{Plots by Set of Benchmarks}
\begin{figure}[H]
  \hspace*{8mm}
    \includegraphics[height=0.9\textheight]{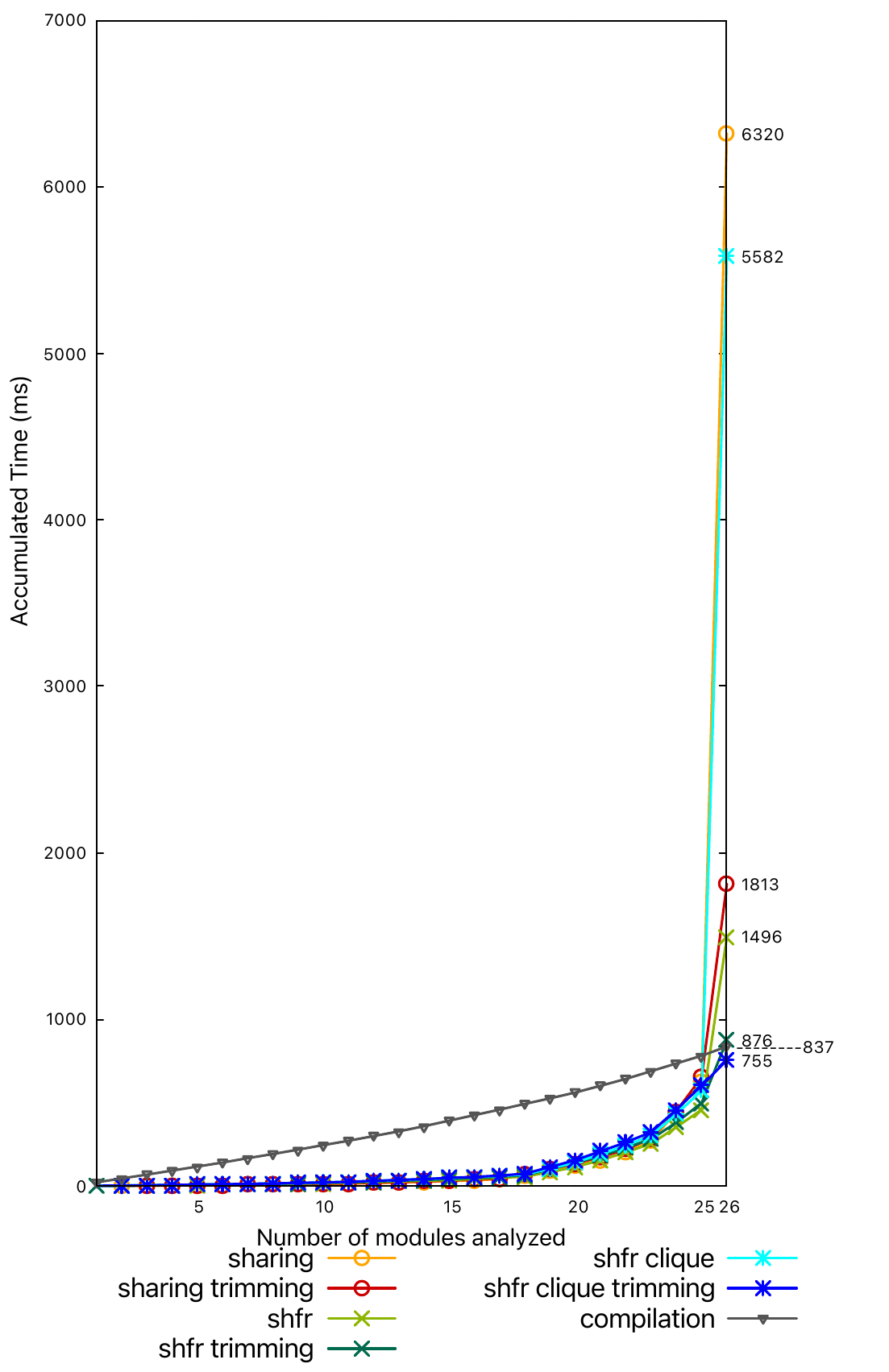}
\caption{Cactus plots aggregating
results of analyzing the classic benchmarks set. \label{figure:cactus-benchs-classic}}
\end{figure}

\newpage
\begin{figure}[H]
  \hspace*{-1mm}
  \includegraphics[height=1\textheight]{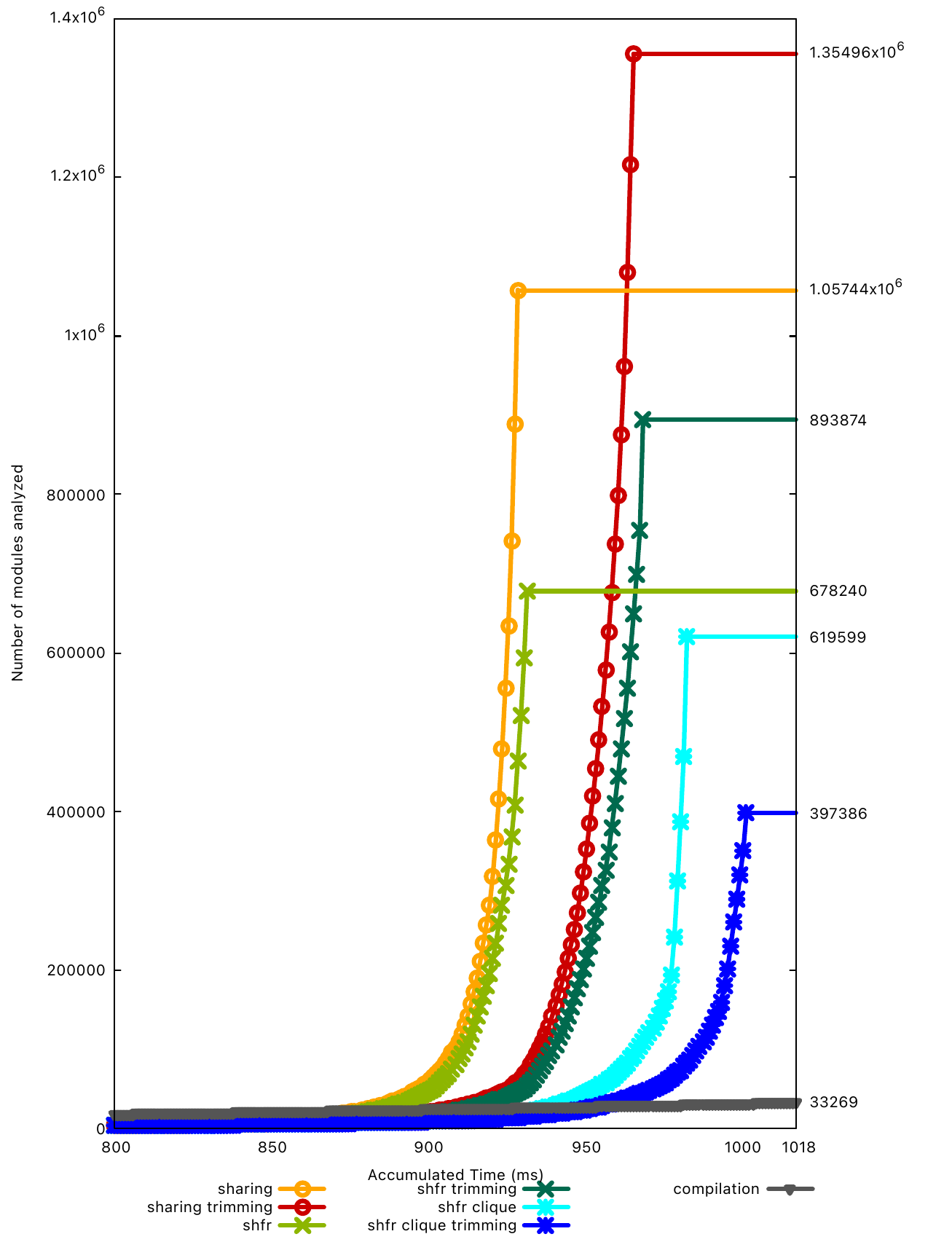}
  \caption{Cactus plots aggregating 
    results of analyzing all 
    libraries in the \ciao system. \label{figure:cactus-benchs-ciaolibs}}
\end{figure}

\newpage
\begin{figure}[H]
  \includegraphics[height=1\textheight]{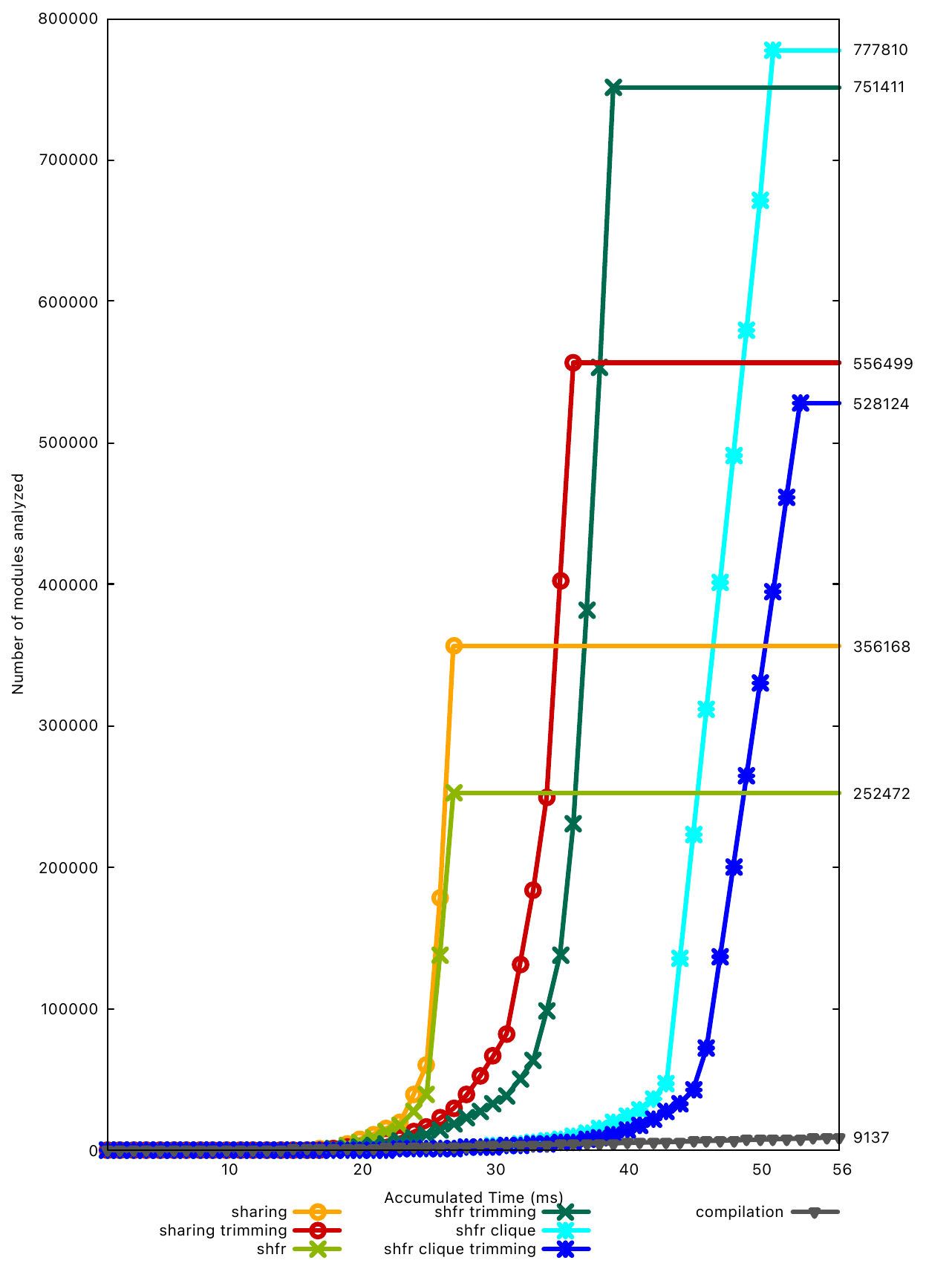}
\caption{Cactus plots aggregating
results of analyzing the source code of \plai. \label{figure:cactus-benchs-plai}}
\end{figure}

\newpage
\begin{figure}[H]
  \hspace*{-9mm}
  \includegraphics[height=1\textheight]{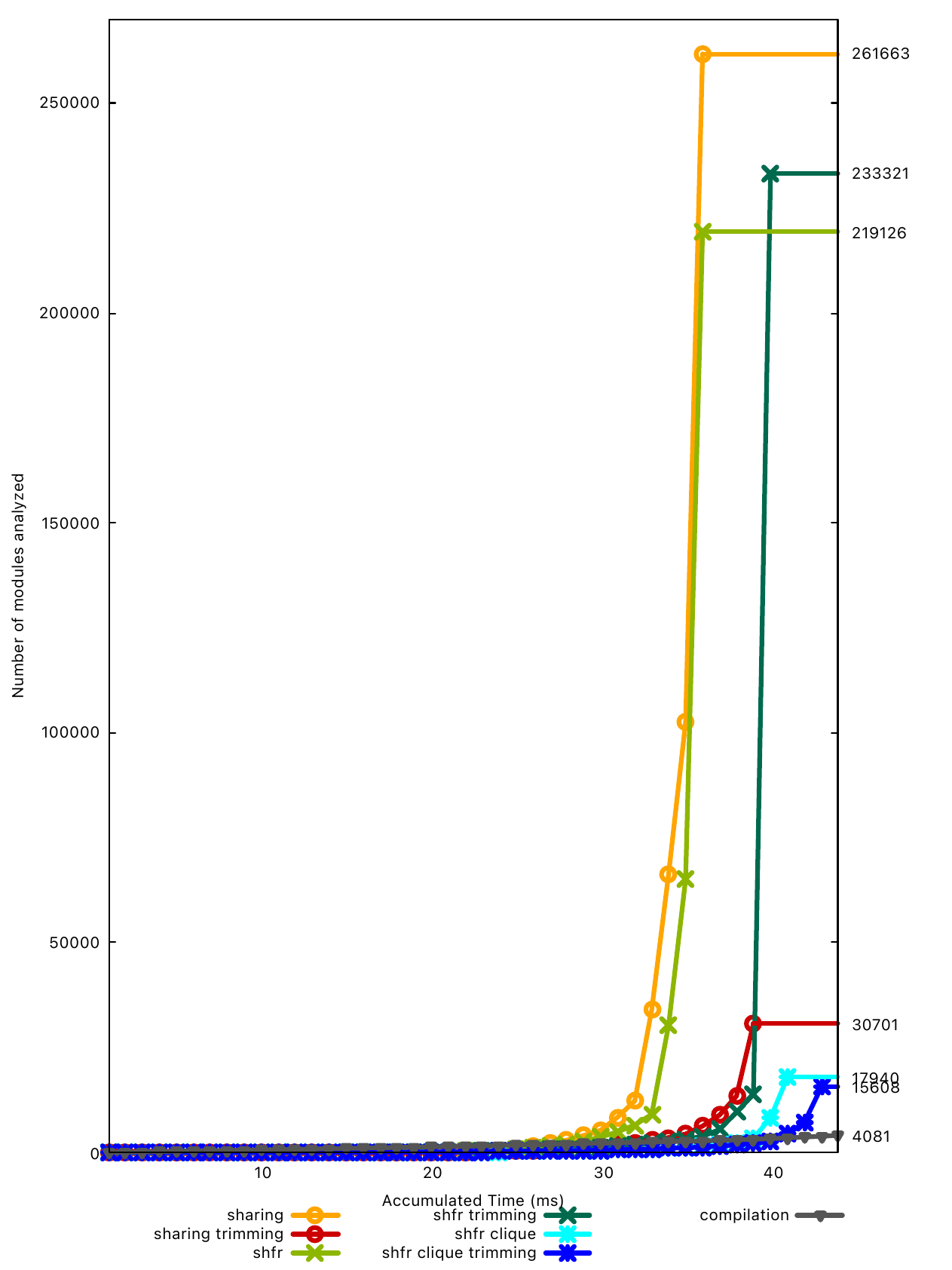}
\caption{Cactus plots aggregating
results of analyzing the source code of s(CASP). \label{figure:cactus-benchs-scasp}}
\end{figure}

\newpage
\begin{figure}[H]
  \includegraphics[height=1\textheight]{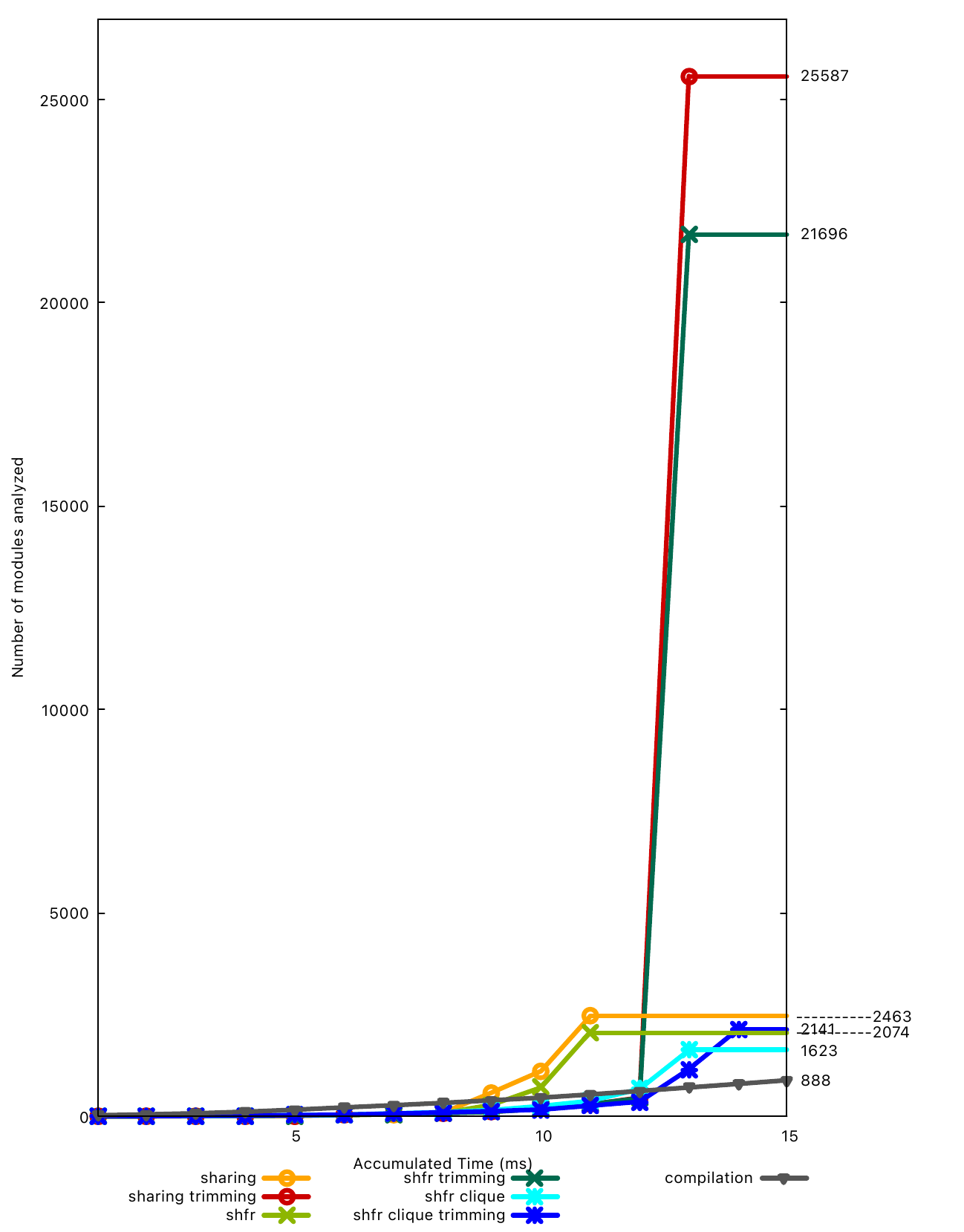}
\caption{Cactus plots aggregating
results of analyzing the source code of Spectector. \label{figure:cactus-benchs-spect}}
\end{figure}

\section{Analysis Times by Sets of Benchmarks.\label{app:tables}}

This section contains all the analysis times obtained while running
the experiments described in Section~\ref{sec:evaluation}. The 
\textbf{TC} columns show the time that the classical analysis requires to
analyze the modules. The \textbf{TT} columns show the time required by
abstract trimming, and \textbf{$\rho$T} shows the relative speedup
given by the formula $\rho$T = TC/TT.

At the end of each table, some statistics are collected.
The number of timeouts with the classic approach
for each analysis, as well as the timeouts obtained while applying
abstract trimming are shown. Then, some statistics related to the
speedups are gathered. First, the mean speedup is provided. Then, we
provide the mean speedups considering only programs which take more
than a time $T$ to be analyzed by the slowest domain (\txt{share}).

We show the number of modules of each kind that are analyzed by each domain. For
example, a row starting with ``$\mu$\textbf{T} > 1s (44-47-101)'' shows
the mean speedup for the benchmarks successfully analyzed such that
the time of analyzing them with \txt{share} is greater than 1 second
or \txt{share} times out. It also shows that in the case of
\txt{share}, 44 of those modules are analyzed successfully with both
approaches, 47 in the case of \txt{shfr}, and 101 in the case of
\txt{shfr-clique}. Notice that both of the approaches
(classic and trimming) need to succeed in order to
compute the means.

These results support the conclusion that the speedups obtained are
greater when the benchmarks are harder, and also that in most
cases, the benchmarks are computed in less than 0.1 seconds.

Finally, please note that, even though some module names may appear
multiple times, this 
does not mean that they are duplicated benchmarks, but rather files with the
same name in different folders and whose code is 
typically different. To make the comparisons, the full paths
where the modules being analyzed are placed are considered; however in
the final table we only present the module names for
readability.

\clearpage
\begin{landscape}
  \subsection{Classic Benchmarks\label{app:class}}
\begin{longtable}{||l||r|r|r|r|p{10mm}||r|r|p{10mm}|r|p{10mm}||r|r|r|r|r||}
\caption{Analysis times and statistics for analyzing the classic benchmarks set.\label{tab:benchs}}\\
\toprule
\hline\hline
    & \multicolumn{5}{l||}{share} & \multicolumn{5}{l||}{shfr} & \multicolumn{5}{l||}{shfr-clique} \\ \hline
\textbf{M}     & \textbf{TC}   & \textbf{TT} & \textbf{$\rho$T} & \textbf{TR} & \textbf{\textbf{$\rho$R}}   & \textbf{TC}            & \textbf{TT}           &  \textbf{$\rho$T}  & \textbf{TR}     & \textbf{$\rho$R}   & \textbf{TC}            & \textbf{TT}           & \textbf{$\rho$T} & \textbf{TR}  & \textbf{$\rho$R}   \\ \hline \hline
\footnotesize
query          & 2.46    & 3.50    & 0.70 & 4.20   & 0.59  & 2.40    & 2.63   & 0.91 & 4.30   & 0.56  & 3.82    & 3.65   & 1.05  & 5.48     & 0.70  \\ 
zebra          & 0.89    & 4.17    & 0.21 & 76.20  & 0.01  & 4.01    & 4.59   & 0.87 & 78.41  & 0.05  & 6.48    & 6.71   & 0.97  & 19848.91 & 0.00  \\ 
boyer          & 42.04   & 44.91   & 0.94 & 49.82  & 0.84  & 47.12   & 50.96  & 0.92 & 53.10  & 0.89  & 38.56   & 40.09  & 0.96  & 44.30    & 0.87  \\ 
tak            & 1.37    & 1.63    & 0.84 & 3.19   & 0.43  & 1.45    & 1.83   & 0.79 & 3.56   & 0.41  & 1.95    & 2.08   & 0.94  & 4.24     & 0.46  \\ 
witt           & 38.81   & 40.18   & 0.97 & 68.53  & 0.57  & 41.88   & 40.19  & 1.04 & 68.53  & 0.61  & 53.54   & 53.10  & 1.01  & 84.32    & 0.63  \\
\hline
\end{longtable}
\normalsize
\end{landscape}

\clearpage
\begin{landscape}
\subsection{  Source code of the \plai Analyzer.\label{app:plai-times}}
\begin{small}
\vspace*{-5mm}
\begin{longtable}{||l||r|r|r||r|r|r||r|r|r||}
  \caption{Analysis times and statistics for analyzing the 
    source code of \plai.\label{tab:plai-times}}\\
\toprule
  \hline\hline
    & \multicolumn{3}{|l|}{share} & \multicolumn{3}{|l|}{shfr} & \multicolumn{3}{|l|}{shfr-clique} \\ \hline
\textbf{module} &  \textbf{TC} & \textbf{TT} & \textbf{$\rho$T}  &  \textbf{TC} & \textbf{TT} & \textbf{$\rho$T} &  \textbf{TC} & \textbf{TT} & \textbf{$\rho$T} \\ \hline \hline
\endfirsthead
\midrule
    & \multicolumn{3}{|l|}{share} & \multicolumn{3}{|l|}{shfr} & \multicolumn{3}{|l|}{shfr-clique} \\ \hline
\textbf{module} &  \textbf{TC} & \textbf{TT} & \textbf{$\rho$T}  &  \textbf{TC} & \textbf{TT} & \textbf{$\rho$T} &  \textbf{TC} & \textbf{TT} & \textbf{$\rho$T} \\ \hline \hline
\endhead
\footnotesize
fixpo\_di & timeout & timeout  & - &  timeout & timeout  & -& 88817 & 66530  & 1.33 \\ 
domains & timeout & 49167  & - &  timeout & 34882  & -& 5063 & 5203  & 0.97 \\ 
intermod & 21486 & 14406  & 1.49 &  11839 & 5512  & 2.15& 10968 & 9839  & 1.11 \\ 
fixpo\_dx\_check\_common & 0.19 & 0.02  & 11.63 &  0.16 & 0.18  & 0.87& 0.18 & 0.18  & 1.00 \\ 
intermod\_punit & timeout & 13342  & - &  timeout & 11593  & -& 775.79 & 459.35  & 1.69 \\ 
trace\_fixp & 188.43 & 137.64  & 1.37 &  164.88 & 122.81  & 1.34& 37.86 & 71.38  & 0.53 \\ 
intermod\_options & 0.25 & 0.32  & 0.78 &  0.29 & 0.27  & 1.07& 0.00 & 0.28  & 0.36 \\ 
domains\_hooks & 4063 & 4079  & 1.00 &  4055 & 4091  & 0.99& 5.72 & 5.68  & 1.01 \\
\hline
\end{longtable}
\end{small}
\end{landscape}
\begin{landscape}
\subsection{ Libraries in the \ciao system.\label{app:libs}}
\begin{small}
\vspace*{-3mm}
\begin{longtable}{||l||r|r|r||r|r|r||r|r|r||}
  \caption{Analysis times and statistics for analyzing the \ciao
    libraries.\label{tab:times-libs}\\[-3mm] }\\
\toprule
  \hline\hline
  & \multicolumn{3}{|l|}{share} &
  \multicolumn{3}{|l|}{shfr} & \multicolumn{3}{|l|}{shfr-clique} \\
  \hline \textbf{module} & \textbf{TC} & \textbf{TT} &
  \textbf{$\rho$T} & \textbf{TC} & \textbf{TT} & \textbf{$\rho$T} &
  \textbf{TC} & \textbf{TT} & \textbf{$\rho$T} \\ \hline \hline
\endfirsthead
\midrule
  & \multicolumn{3}{|l|}{share} &
  \multicolumn{3}{|l|}{shfr} & \multicolumn{3}{|l|}{shfr-clique} \\
  \hline \textbf{module} & \textbf{TC} & \textbf{TT} &
  \textbf{$\rho$T} & \textbf{TC} & \textbf{TT} & \textbf{$\rho$T} &
  \textbf{TC} & \textbf{TT} & \textbf{$\rho$T} \\ \hline \hline
\endhead
\footnotesize
  initial & 0.25 & 0.24 &
  1.05 & 0.25 & 0.35 &
  0.70& 0.29 & 0.42 &
  0.69 \\   
  operators &
  16.06 & 17.93 & 0.90 &
  16.60 & 13.70 & 1.21&
  14.58 & 19.98 & 0.73 \\
  
   unittestdecls & 0.25 &
  0.28 & 0.92 & 0.24 &
  0.33 & 0.74& 0.25 &
  0.26 & 0.96 \\ 
  system & 2795.54 &
  2710.89 & 1.03 &
  3288.71 & 3316.89 &
  0.99& 4436.77 & 4552.79
  & 0.97 \\ 
   default &
  0.26 & 0.47 & 0.55 &
  0.26 & 0.24 & 1.06&
  0.26 & 0.30 & 0.88 \\
  
   stream\_utils & 18.21 &
  13.45 & 1.35 & 19.93 &
  17.21 & 1.16& 24.32 &
  24.71 & 0.98 \\ 
  dynmod\_holder & 0.63 &
  0.63 & 0.99 & 0.71 &
  0.67 & 1.06& 0.69 &
  1.01 & 0.69 \\ 
  dict & 3534.96 &
  3545.48 & 1.00 &
  1819.39 & 1867.04 &
  0.97& 171.06 & 194.10 &
  0.88 \\
\hline
\end{longtable}
\end{small}
\end{landscape}

\begin{landscape}
  \subsection{Source code of the s(CASP) System\label{app:scasp}}
  \begin{longtable}{||l||r|r|r||r|r|r||r|r|r||}

\caption{Analysis times and statistics for analyzing the source code
  of the s(CASP) system.\label{tab:scasp}}\\
\hline\hline
\toprule
  \hline\hline
    & \multicolumn{3}{|l|}{share} & \multicolumn{3}{|l|}{shfr} & \multicolumn{3}{|l|}{shfr-clique} \\ \hline
\textbf{module} &  \textbf{TC} & \textbf{TT} & \textbf{$\rho$T}  &  \textbf{TC} & \textbf{TT} & \textbf{$\rho$T} &  \textbf{TC} & \textbf{TT} & \textbf{$\rho$T} \\ \hline \hline
\endfirsthead
\midrule
    & \multicolumn{3}{|l|}{share} & \multicolumn{3}{|l|}{shfr} & \multicolumn{3}{|l|}{shfr-clique} \\ \hline
\textbf{module} &  \textbf{TC} & \textbf{TT} & \textbf{$\rho$T}  &  \textbf{TC} & \textbf{TT} & \textbf{$\rho$T} &  \textbf{TC} & \textbf{TT} & \textbf{$\rho$T} \\ \hline \hline
\endhead
\footnotesize
scasp\_load\_compiled & 7.82 & 9.24  & 0.85 &  8.85 & 10.00  & 0.89& 11.32 & 12.50  & 0.91 \\ 
clp\_call\_stack & 4.97 & 5.51  & 0.90 &  5.56 & 5.51  & 1.01& 7.36 & 7.17  & 1.03 \\ 
test\_results\_c\_forall & 0.14 & 0.20  & 0.73 &  0.15 & 0.22  & 0.68& 0.15 & 0.17  & 0.87 \\ 
test\_results & 0.17 & 0.14  & 1.22 &  0.13 & 0.18  & 0.73& 0.18 & 0.14  & 1.30 \\ 
clp\_disequality\_rt & 15.29 & 12.35  & 1.24 &  17.29 & 17.47  & 0.99& 14.24 & 18.99  & 0.75 \\ 
clp\_disequality & 0.29 & 0.24  & 1.20 &  0.31 & 0.29  & 1.08& 0.25 & 0.27  & 0.94 \\ 
clp\_disequality\_test & 7.37 & 8.86  & 0.83 &  8.32 & 9.91  & 0.84& 6.89 & 10.91  & 0.63 \\ 
scasp & timeout & timeout  & - &  timeout & timeout  & -& timeout & 8419.57  & - \\
\end{longtable}

\end{landscape}
\clearpage
\begin{landscape}
  \subsection{Source Code of the Spectector Tool\label{app:spect}}
  \begin{longtable}{||l||r|r|r||r|r|r||r|r|r||}
\caption{Analysis times and statistics for analyzing the source code
  of the Spectector tool.\label{tab:spect}}\\
\hline\hline
\toprule
  \hline\hline
    & \multicolumn{3}{|l|}{share} & \multicolumn{3}{|l|}{shfr} & \multicolumn{3}{|l|}{shfr-clique} \\ \hline
\textbf{module} &  \textbf{TC} & \textbf{TT} & \textbf{$\rho$T}  &  \textbf{TC} & \textbf{TT} & \textbf{$\rho$T} &  \textbf{TC} & \textbf{TT} & \textbf{$\rho$T} \\ \hline \hline
\endfirsthead
\midrule
    & \multicolumn{3}{|l|}{share} & \multicolumn{3}{|l|}{shfr} & \multicolumn{3}{|l|}{shfr-clique} \\ \hline
\textbf{module} &  \textbf{TC} & \textbf{TT} & \textbf{$\rho$T}  &  \textbf{TC} & \textbf{TT} & \textbf{$\rho$T} &  \textbf{TC} & \textbf{TT} & \textbf{$\rho$T} \\ \hline \hline
\endhead
\footnotesize
spectector\_flags & 3.71 & 3.96  & 0.94 &  4.08 & 4.36  & 0.94& 4.87 & 4.73  & 1.03 \\ 
muasm\_semantics & 1351.25 & 164.39  & 8.22 &  1364.86 & 150.69  & 9.06& 98.89 & 94.20  & 1.05 \\ 
muasm\_print & 25.04 & 26.79  & 0.93 &  13.15 & 15.23  & 0.86& 13.44 & 10.91  & 1.23 \\ 
muasm\_syntax & 0.25 & 0.25  & 0.99 &  0.27 & 0.25  & 1.07& 0.31 & 0.27  & 1.18 \\ 
spectector\_stats & timeout & 71.00  & - &  timeout & 112.20  & -& 292.50 & 29.88  & 9.79 \\ 
spectector & timeout & timeout  & - &  timeout & timeout  & -& timeout & timeout  & - \\ 
muasm\_program & 1.41 & 5.98  & 0.24 &  5.04 & 5.54  & 0.91& 8.40 & 9.36  & 0.90 \\ 
spectector\_noninter & timeout & 25132.10  & - &  timeout & 21251.12  & -& timeout & 801.15  & - \\
\end{longtable}
\end{landscape}

\clearpage{}%

\end{document}